\documentclass[journal]{IEEEtran}
\usepackage[T1]{fontenc}
\usepackage[utf8]{inputenc}
\usepackage{cite}
\usepackage{amsmath,amssymb,amsfonts,bm}
\usepackage{amsthm}
\usepackage{graphicx}
\usepackage{booktabs}
\usepackage{array}
\usepackage{url}
\usepackage{stfloats}       
\usepackage{tikz}
\usetikzlibrary{arrows.meta,positioning,fit,calc}
\usepackage[caption=false,font=footnotesize]{subfig}
\usepackage{comment}
\usepackage[nolist,nohyperlinks]{acronym}
\usepackage{balance}

\usepackage[switch]{lineno}
\usepackage[named]{algo}
\usepackage{algpseudocode}
\usepackage{algorithm}
\algnewcommand{\LineComment}[1]{\State \(\triangleright\) #1}
\algtext*{EndFor}
\algtext*{EndIf}
\algtext*{EndWhile}
\algtext*{EndFunction}

\graphicspath{{./plts/}{./figs/}}

\newcolumntype{L}[1]{>{\raggedright\arraybackslash}p{#1}}

\begin{document}
\begin{acronym}[GENERAL]
    \acro{CVQKD}[CV-QKD]{continuous-variable quantum key distribution}
    \acro{DSP}{digital signal processing}
    \acro{FSO}{free-space optical}
    \acro{IID}{independent and identically distributed}
    \acro{IQR}{interquartile range}
    \acro{KL}{Kullback--Leibler}
    \acro{LMMSE}{linear minimum mean-square error}
    \acro{MMSE}{minimum mean-square error}
    \acro{LS}{least squares}
    \acro{MIMO}{multiple-input multiple-output}
    \acro{MSFE}{mean squared Frobenius error}
    \acro{NRMSE}{normalized root mean squared error}
    \acro{ReLU}{rectified linear unit}
    \acro{SATCOM}{satellite communications}
    \acro{VAE}{variational autoencoder}
\end{acronym}

\title{Physical Prior Mismatch in Variational Dual-Polarization Quantum SATCOM Channel Estimation}

\author{Anders~M.~Buvarp,~\IEEEmembership{Senior Member,~IEEE,}
        Stefan~Werner,~\IEEEmembership{Fellow,~IEEE}
\thanks{A. M. Buvarp and S. Werner are with the Department of Electronic Systems, Norwegian University of Science and Technology, Trondheim, Norway. E-mail: anders.buvarp@ntnu.no. Stefan Werner is also with the Department of Information and Communications Engineering, Aalto University, 00076, Finland.}}

\maketitle

\begin{abstract}
Dual-polarization quantum satellite communications require accurate estimation of complex polarization channels. We investigate physical latent-prior mismatch in variational estimation of a $2\times2$ complex channel, using eight Cartesian latent coordinates and a fixed latent-to-channel mapping. Gaussian, matched and sharpened Rician, Laplace, and Cauchy priors are compared under a fixed neural architecture and forward Kullback--Leibler objective. Models are trained for $300{,}000$ updates with $\zeta=1/(2N_{\mathrm{est}})$ and evaluated using three training seeds and $10{,}000$ shared held-out blocks per operating point. The Gaussian and Rician priors coincide at the Rayleigh limit. At higher Rician factors, the matched prior improves estimation accuracy, especially with fewer pilots. At $K=100$ and $N_{\mathrm{est}}=32$, the matched Rician variational estimator reduces mean squared Frobenius error by $2.50\%$ relative to linear minimum mean-square error (LMMSE). We also derive a joint Rician Bayesian posterior-mean estimator that incorporates the complete least-squares noise covariance and reduces posterior integration to one-dimensional quadrature. It improves on LMMSE by $3.33\%$ at the same operating point and achieves lower observed error than the scalar Rician and both neural estimators at all 12 operating points. Additional Rician prior sharpening increases error, while Laplace and Cauchy controls remain near the Gaussian baseline.
\end{abstract}


\section{Introduction}

Dual-polarization coherent receivers estimate a complex
$2\times2$ channel to compensate for polarization mixing,
amplitude variations, and phase offsets
\cite{Pereira2023,Almeida2025,Guilherme2025,Almeida2025a}.
In free-space optical and quantum satellite communication
systems, propagation and receiver impairments make accurate
channel estimation an important component of receiver
processing \cite{Costa2024,Muga2024,Jaiswal2018,Kim2026}.
Pilot-aided estimation provides a setting in which the
relationship between channel observations and statistical
prior information can be examined directly.

Variational channel estimators incorporate prior information
through regularization of an approximate posterior
distribution. When the latent coordinates represent physical
channel coefficients, the assumed prior specifies a
statistical model of those coefficients. Consequently,
a conventional isotropic Gaussian prior can introduce
mismatch when the physical channel follows a non-Gaussian
distribution. The estimation consequences of this mismatch
are the focus of the present work.

Variational inference provides a probabilistic alternative to deterministic
channel estimators by representing the unknown channel through a latent random
variable~\cite{Kingma2014}. Variational methods have subsequently been applied
to channel estimation and receiver inference in communication systems
\cite{Tang2022,Fesl2024}. In a conventional variational formulation, an
encoder parameterizes an approximate posterior
$q_\phi(\mathbf{z}|\mathbf{s})$, where $\mathbf{s}$ denotes
channel-relevant receiver statistics, and training balances a reconstruction
term against the forward \ac{KL} divergence. The prior
$p(\mathbf{z})$ therefore enters directly into the estimator through the
regularization term and can influence the inferred channel state whenever the
\ac{KL} contribution is non-negligible.

The standard isotropic Gaussian prior is convenient and well matched when the
chosen latent coordinates are zero-mean Gaussian with identity covariance.
In physical channel estimation, however, the latent variable may represent
the channel coefficients themselves. The prior then becomes a statistical
assumption about the physical channel rather than only a latent-space
regularization choice. When the channel law departs from the isotropic Gaussian
model, the estimator is regularized toward a distribution that does not match
the physical parameter being inferred.

The central issue is whether the conventional isotropic Gaussian prior remains
an adequate approximation across the relevant channel regimes, or whether
mismatch between the assumed latent prior and the physical channel
distribution produces measurable degradation in estimation performance.

The role of the prior is therefore twofold. It represents an assumption about
the physical channel distribution, but it also acts as a regularizer within
the variational objective. Consequently, the estimation performance depends
on both the assumed prior and its contribution to the training objective.
In this work, we compare physically matched and mismatched priors under
a fixed variational formulation to determine when channel-specific prior
information improves estimation accuracy.

To the best of our knowledge, prior work has not examined
physical latent-prior mismatch under a fixed latent-to-channel
mapping while holding the posterior family, forward \ac{KL}
objective, observation model, and estimator architecture fixed.
We study this question for dual-polarization quantum
\ac{SATCOM} channel estimation using a $2\times2$ complex
polarization channel. Its four complex coefficients define an
eight-dimensional real latent variable through their normalized
real and imaginary components. The latent-to-channel mapping
is fixed by this physical parameterization, eliminating
learned-decoder flexibility from the prior comparison.
The estimator uses the \ac{LS} channel estimate and normalized
pilot Gram matrix as input to a diagonal-Gaussian approximate
posterior.

The channel coefficients follow independent random-phase
Rician distributions, providing a controlled receiver-side
estimation model rather than a general statistical model of
satellite optical polarization channels. We compare the
conventional isotropic Gaussian prior with the exact Rician
latent prior induced by this channel law. At $K{=}0$, the
two priors coincide analytically. Increasing $K$ introduces
progressively stronger prior mismatch without changing the
estimator architecture or observation model. We evaluate
$K\in\{10,30,100\}$ over multiple pilot-block lengths using
paired Monte Carlo trials with identical channel and data
realizations across estimators.

The matched Rician prior improves channel \ac{NRMSE} relative
to the Gaussian prior, with the largest gains at high $K$ and
small $N_{\mathrm{est}}$. At $K=100$ and $N_{\mathrm{est}}=32$,
the matched Rician \ac{VAE} reduces \ac{MSFE} by $2.50\%$
relative to \ac{LMMSE}. The joint Rician \ac{MMSE}
estimator achieves a $3.33\%$ reduction at the same
operating point and lower observed \ac{MSFE} than the
scalar Rician posterior-mean estimator and both neural
estimators at all $12$ tested operating points. Sensitivity
experiments show that additional Rician prior sharpening
increases estimation error, while the tested Laplace and
Cauchy controls do not reproduce the matched Rician gains.

The joint Rician \ac{MMSE} estimator exploits the known
factorized Rician channel law and the complete \ac{LS}
noise covariance. Analytical integration of the Gaussian
component and one Rician phase reduces the posterior
calculation to one-dimensional quadrature. In contrast,
the variational estimator provides a controlled setting
for studying physical prior mismatch. The observed
performance gap also motivates extending the analysis
to channel models for which model-matched posterior-mean
inference is computationally less tractable.

The main contributions of this work are as follows:\\
\emph{1) Dual-polarization channel estimator:} We construct a variational dual-polarization channel estimator in
which the latent variables correspond directly to the physical complex
channel coefficients and the latent-to-channel mapping is fixed, providing
a controlled setting for studying latent-prior mismatch.\\
\emph{2) Physically matched Rician prior:} We examine mismatch between the assumed latent prior and the physical
channel distribution by comparing the conventional isotropic Gaussian prior
with the physically matched Rician prior induced by the simulated channel
law while holding the posterior family, forward \ac{KL} objective, observation
model, and estimator architecture fixed.\\
\emph{3) Joint Rician MMSE estimator:} We derive a
model-matched Bayesian posterior-mean estimator that accounts
for the complete complex \ac{LS} noise covariance. By
analytically integrating the Gaussian component and one
Rician phase, we reduce the posterior integration to
one-dimensional quadrature with complexity $O(Q)$,
providing an optimal Bayesian reference under the
assumed channel and observation models.\\
\emph{4) Paired Monte Carlo evaluation:} We characterize
the dependence of prior mismatch on Rician line-of-sight
strength and channel estimation block length using paired
Monte Carlo statistics. The matched Rician \ac{VAE}
achieves an \ac{MSFE} improvement of up to $2.50\%$
over \ac{LMMSE}, while the joint Rician \ac{MMSE}
estimator achieves $3.33\%$ at the same operating point.
The joint estimator gives lower observed \ac{MSFE}
than the scalar Rician estimator and both neural
estimators at all $12$ operating points.\\
\emph{5) Prior mismatch and non-Gaussian controls:} We examine
sensitivity to the assumed Rician factor and compare the matched
Rician prior with Gaussian, Laplace, and Cauchy alternatives.
The physically matched Rician prior gives the lowest estimation
error among the tested priors, whereas additional Rician prior
sharpening generally increases the error.  The Laplace and Cauchy controls remain close to the Gaussian baseline and do not reproduce the matched Rician gains under the tested conditions.

The objective is to identify the physical channel regimes
in which the isotropic Gaussian prior remains an
adequate approximation. The
results show that the physically matched Rician prior can improve estimation accuracy over Gaussian inference, particularly in pilot-limited, high-$K$ regimes, while additional prior sharpening does not improve performance
under the tested conditions. The reported results are receiver-level results. Protocol-level \ac{CVQKD} security analysis is outside the scope of this paper.

The remainder of this paper is organized as follows:
Section~\ref{sec:bg} reviews related work.
Section~\ref{sec:model} introduces the dual-polarization channel model,
estimation statistics, and physical latent parameterization.
Section~\ref{sec:vae} presents the variational channel estimator.
Section~\ref{sec:prior} defines the Gaussian, Rician, and non-Gaussian
control priors.
Section~\ref{sec:training} describes the training and estimation procedure.
Section~\ref{sec:simulations} defines the simulation benchmark and paired
Monte Carlo protocol.
Section~\ref{sec:results} presents the numerical results.
Section~\ref{sec:disc} discusses latent-prior mismatch,
the role of physical prior information, the limitations
of the adopted channel and estimation models, and computational complexity.
Section~\ref{sec:conclusion} concludes the paper.

\noindent \emph{Notation:}
Bold lowercase and uppercase letters denote vectors and matrices, respectively.
The real and complex fields are denoted by $\mathbb{R}$ and $\mathbb{C}$.
The Hermitian transpose and complex conjugate are denoted by $(\cdot)^H$
and $\overline{(\cdot)}$, respectively. The Euclidean and Frobenius norms are
denoted by $\lVert\cdot\rVert_2$ and $\lVert\cdot\rVert_F$, and the identity
matrix of size $n$ is denoted by $\bm I_n$.
At discrete time $t$, $\bm x_t,\bm y_t,\bm n_t\in\mathbb{C}^2$ denote the
transmitted, received, and additive-noise vectors. The effective
dual-polarization channel is $\bm H\in\mathbb{C}^{2\times2}$ and is assumed
constant over an estimation block. The corresponding transmitted and received
sample matrices are denoted by $\bm X$ and $\bm Y$. Estimated quantities are
marked with hats. The latent channel vector is denoted by
$\bm z\in\mathbb{R}^8$. The approximate posterior is
$q_\phi(\bm z\mid\bm s)$, and the assumed latent prior is denoted by
$p(\bm z)$. The conventional Gaussian prior is written as
$p_{\mathrm G}(\bm z)$, the Rician prior as $p_K(\bm z)$, and the Laplace
and Cauchy control priors as $p_{\mathrm L}(\bm z)$ and
$p_{\mathrm C}(\bm z)$, respectively. The \ac{KL} divergence is
denoted by $D_{\mathrm{KL}}(Q\Vert P)$. The real Gaussian distribution
with mean $\bm\mu$ and covariance $\bm\Sigma$ is denoted by
$\mathcal{N}(\bm\mu,\bm\Sigma)$, while
$\mathcal{N}_{\mathrm C}(\bm\mu,\bm\Sigma)$ denotes a proper complex Gaussian
distribution. Bachmann--Landau notation is denoted by $\mathcal{O}(\cdot)$.

\section{Related Work}
\label{sec:bg}

\Ac{CVQKD} provides a relevant setting for dual-polarization
free-space links using coherent detection and heterodyne
measurements. Satellite-to-ground feasibility has been
investigated using realistic channel models \cite{Dequal2021},
while terrestrial free-space \ac{CVQKD} has been demonstrated
under high background noise \cite{Zheng2025}.
Ruppert \emph{et al.}~\cite{Ruppert2019} studied fading
channel estimation for free-space quantum communication.
The present work addresses receiver-side estimation rather
than protocol-level security analysis.

Dual-polarization optical receivers compensate for
polarization drift, imbalance, and mixing using digital
signal processing \cite{Pereira2023,Almeida2025,
Guilherme2025,Almeida2025a}.
Conventional approaches include adaptive complex MIMO
equalization and polarization-control algorithms
\cite{Savory2010,Noe2010,Roudas2010,Johannisson2011}.
Unlike blind equalization, pilot-aided estimation exploits
known transmitted symbols to estimate the physical channel.

Variational inference represents unknown channel or signal
quantities through latent variables and an approximate
posterior regularized toward an assumed prior
\cite{Kingma2014}. Applications include unsupervised
equalization and decoding \cite{Caciularu2020}, blind MIMO
channel estimation \cite{Tang2022}, and coherent optical
equalization \cite{Lauinger2022}. Qin \emph{et al.}
investigated vector-quantized \ac{VAE} equalization for
free-space optical MIMO transmission \cite{Qin2025}
and subsequently incorporated nonlinear MIMO filtering
and receiver-impairment compensation \cite{Qin2026}.
These studies primarily address receiver inference or
equalization rather than physical latent-prior mismatch
under a fixed forward \ac{KL} objective.

Physical channel structure has also been incorporated
into variational estimators. Guo \emph{et al.} used
physical parameters as latent variables with a fixed
propagation-model decoder \cite{Guo2025}.
Structured priors have been investigated for massive MIMO
\cite{Cheng2020}, RIS-assisted mmWave channels
\cite{Nakul2024}, and IRS-assisted sparse channel
estimation \cite{Yang2025}. Thus, neither physical
latent parameterization nor non-Gaussian channel priors
constitutes a contribution by itself.

Other studies learn channel distributions for Bayesian
estimation. Baur \emph{et al.} used variational models
to approximate channel statistics for \ac{MMSE}
estimation \cite{Baur2022,Baur2024}.
Fesl \emph{et al.} investigated conditionally Gaussian
latent models \cite{Fesl2024} and generative
Gaussian-mixture priors for quantized channel estimation
\cite{Fesl2024Generative}. These methods use learned
or structured distributions for estimation, whereas
PRIOR compares prescribed physical and mismatched
latent priors within the same variational estimator.

Cremer \emph{et al.}~\cite{Cremer2018} distinguish
posterior approximation error from amortization error
in variational inference. In contrast, the Bayesian
\ac{MMSE} estimator minimizes the expected squared
estimation error under the assumed channel and
observation models. In this work, we derive the joint
Rician posterior mean by accounting for the complete
\ac{LS} noise covariance and analytically reducing the
posterior integration to one dimension. This provides
a model-matched Bayesian reference for evaluating
the estimation gap of the neural variational estimators.

In our related work \cite{Anders2026}, we investigate
variational blind equalization for dual-polarization
free-space quantum communications, emphasizing the
choice of divergence measure. The present study instead
examines physical latent-prior mismatch in pilot-aided
channel estimation under a fixed forward \ac{KL}
objective.

\begin{figure}[t!]
\centering
\begin{tikzpicture}[
    node distance=5mm,
    >=Latex,
    block/.style={
        draw,
        rounded corners=2pt,
        align=center,
        minimum height=8mm,
        text width=0.82\columnwidth,
        inner sep=4pt,
        font=\footnotesize
    },
    arrow/.style={
        ->,
        thick
    }
]

\node[block] (tx) {
    Dual-polarization transmitter\\
    $\bm x_t=[x_{H,t},\,x_{V,t}]^{T}$
};

\node[block, below=of tx] (channel) {
    Dual-polarization free-space channel\\
    $\bm y_t=\bm H\bm x_t+\bm n_t$
};

\node[block, below=of channel] (rx) {
    Transmitted and received samples $\bm X$ and $\bm Y$
};

\node[block, below=of rx] (stats) {
    Channel statistics\\
    $\bm S_{yx}=\bm Y\bm X^{H}$,
    \qquad
    $\bm S_{xx}=\bm X\bm X^{H}$
};

\node[block, below=of stats] (encoder) {
    Variational channel estimator\\
    $q_{\phi}(\bm z|\bm s)$
};

\node[block, below=of encoder] (prior) {
    Latent prior regularization\\
    Assumed latent prior $p(\bm z)$:\\
    Gaussian, Rician, Laplace, or Cauchy\\
    $\zeta D_{\mathrm{KL}}\!\left(q_{\phi}\Vert p\right)$
};

\node[block, below=of prior] (physical) {
    Fixed physical channel mapping\\
    $\bm z \rightarrow \widehat{\bm H}=\bm H(\bm z)$
};

\node[block, below=of physical] (dsp) {
    Receiver evaluation:\\
Channel NRMSE and MSFE
};

\draw[arrow] (tx) -- (channel);
\draw[arrow] (channel) -- (rx);
\draw[arrow] (rx) -- (stats);
\draw[arrow] (stats) -- (encoder);
\draw[arrow] (encoder) -- (prior);
\draw[arrow] (prior) -- (physical);
\draw[arrow] (physical) -- (dsp);

\end{tikzpicture}
\caption{System model for variational dual-polarization channel estimation.
The estimator forms channel-relevant statistics from transmitted and received
sample blocks and infers a posterior distribution over the normalized Cartesian
channel coordinates. The latent posterior is regularized toward an assumed
prior, including the conventional isotropic Gaussian model, Rician priors, and
the Laplace and Cauchy control priors. A fixed physical mapping converts the
latent channel coordinates to the estimated $2\times2$ complex polarization
channel.}
\label{fig:prior_system_model}
\end{figure}
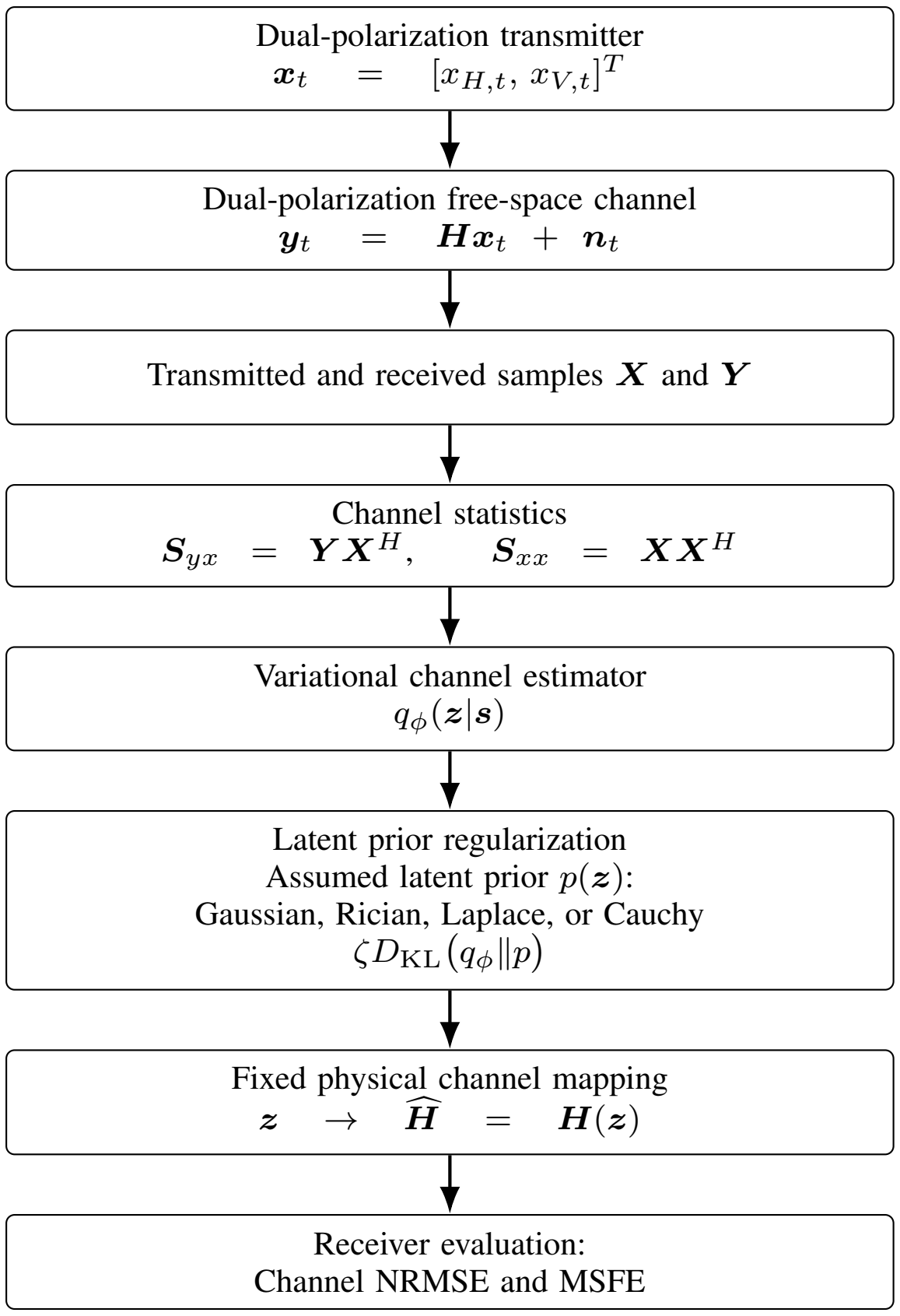

\vspace{-1mm}
\section{Dual-Polarization Free-Space Channel Model}
\label{sec:model}

Fig.~\ref{fig:prior_system_model} summarizes the considered
dual-polarization free-space receiver model and the statistics-based
variational channel estimation procedure. The receiver observes two complex
baseband polarization components whose input--output relation is represented
by a $2\times2$ complex MIMO channel. The objective is to estimate the
effective polarization channel from a finite block of transmitted and received
samples and to obtain a posterior distribution over the corresponding physical
channel parameters.

\subsection{Dual-Polarization Signal Model}

We consider pilot-aided estimation of the effective
dual-polarization channel $\bm H\in\mathbb C^{2\times2}$.
At symbol time $t$, the transmitted and received vectors
satisfy
\begin{equation}
\label{eq:system_model}
    \bm x_t =
    \begin{bmatrix}
        x_{H,t}\\
        x_{V,t}
    \end{bmatrix}
    \in\mathbb C^2,
    \qquad
    \bm y_t=\bm H\bm x_t+\bm n_t,
\end{equation}
where $H$ and $V$ identify the two polarization branches.
The channel matrix is parameterized as
\begin{equation}
    \bm H=
    \begin{bmatrix}
        h_{HH}&h_{HV}\\
        h_{VH}&h_{VV}
    \end{bmatrix}.
\end{equation}
The diagonal and off-diagonal coefficients represent direct
and cross-polarization transfer, respectively, incorporating
effective propagation and receiver effects such as polarization
rotation, differential gain and phase, propagation loss, and
residual calibration mismatch.

All four complex coefficients are estimated from known
transmitted samples. Their statistical distribution is
specified separately in Section~V, allowing the assumed
latent prior to be varied without changing the observation
model.
The receiver noise is modeled as
\begin{equation}
    \bm n_t\sim
    \mathcal N_{\mathrm C}
    (\bm 0,\sigma_n^2\bm I_2).
\end{equation}
The channel remains fixed within each estimation block
and changes independently between blocks.

\subsection{Block Channel Estimation Model}

For a block of $N_{\mathrm{est}}$ transmitted sample vectors, define
\begin{equation}
    \bm X =
    \begin{bmatrix}
        \bm x_1 & \cdots & \bm x_{N_{\mathrm{est}}}
    \end{bmatrix}
    \in\mathbb{C}^{2\times N_{\mathrm{est}}}
\end{equation}
and
\begin{equation}
    \bm Y =
    \begin{bmatrix}
        \bm y_1 & \cdots & \bm y_{N_{\mathrm{est}}}
    \end{bmatrix}
    \in\mathbb{C}^{2\times N_{\mathrm{est}}}.
\end{equation}
The block observation model is
\begin{equation}
    \bm Y = \bm H\bm X + \bm N,
    \label{eq:block_model}
\end{equation}
where $\bm H$ is assumed constant over the estimation block.

Rather than presenting the neural estimator with the raw vectorized
$\bm X$ and $\bm Y$, we form the cross- and auto-statistics
\begin{align}
    \bm S_{yx} &= \bm Y\bm X^{H},\\
    \bm S_{xx} &= \bm X\bm X^{H}.
    \label{eq:channel_statistics}
\end{align}
These matrices provide a compact second-order representation of the
transmitted and received estimation blocks for the variational estimator.

For reference, the corresponding least-squares channel estimate is
\begin{equation}
    \widehat{\bm H}_{\mathrm{LS}}
    =
    \bm Y\bm X^{H}
    \left(
        \bm X\bm X^{H}
    \right)^{-1},
    \label{eq:ls_channel_estimate}
\end{equation}
when $\bm X\bm X^{H}$ is nonsingular. 

For comparison, the analytical \ac{LMMSE} channel estimate is
\begin{equation}
    \widehat{\bm H}_{\mathrm{LMMSE}}
    =
    \bm S_{yx}
    \left(
        \bm S_{xx}+\sigma_n^2\bm I_2
    \right)^{-1}.
    \label{eq:lmmse_channel_estimate}
\end{equation}
The channel coefficients have zero mean and unit variance under the
random-phase Rician model in \eqref{eq:rician_channel}.
The covariance of $\operatorname{vec}(\bm H)$, where
$\operatorname{vec}(\cdot)$ denotes column-wise vectorization,
is therefore $\bm I_4$ for every $K$.
Consequently, the analytical \ac{LMMSE} estimator depends on the
second-order channel statistics but not explicitly on $K$.

The variational estimator uses the \ac{LS} channel estimate and the
normalized pilot Gram matrix $\bm S_{xx}/N_{\mathrm{est}}$
to construct its $16$-dimensional real input. These features
are standardized using training-set statistics and mapped
to an approximate posterior over the physical channel
coordinates.

\subsection{Scalar Rician Posterior-Mean Benchmark}

We also consider a scalar matched Rician posterior-mean estimator
as an analytical reference. For each coefficient of the \ac{LS} channel
estimate, let
\begin{equation}
    r_{ij}=[\widehat{\bm H}_{\mathrm{LS}}]_{ij},
    \qquad
    \tau_j=\sigma_n^2
    [\bm S_{xx}^{-1}]_{jj},
\end{equation}
where $\tau_j$ is the variance of the \ac{LS} estimation error for
transmitter column $j$. Under the random-phase Rician model in
\eqref{eq:rician_channel}, define
\begin{equation}
    a=\sqrt{\frac{K}{K+1}},
    \qquad
    v=\frac{1}{K+1},
    \qquad
    \kappa_{ij}=\frac{2a|r_{ij}|}{v+\tau_j}.
\end{equation}
The scalar posterior-mean estimate is
\begin{equation}
\begin{aligned}
    \widehat h_{ij,\mathrm R}
    ={}&
    \frac{v}{v+\tau_j}r_{ij}\\
    &+
    \frac{a\tau_j}{v+\tau_j}
    \frac{r_{ij}}{|r_{ij}|}
    \frac{I_1(\kappa_{ij})}{I_0(\kappa_{ij})},
\end{aligned}
\label{eq:scalar_rician_posterior_mean}
\end{equation}
where $I_0(\cdot)$ and $I_1(\cdot)$ are modified Bessel functions
of the first kind. At $r_{ij}=0$, the estimate is defined by
continuity as $\widehat h_{ij,\mathrm R}=0$.

This estimator is the exact posterior mean conditioned on an
individual \ac{LS} channel coefficient and the pilot matrix.
It does not account for correlations between \ac{LS} estimation errors
associated with different transmitter columns and is therefore
not the full joint-channel Bayesian \ac{MMSE} estimator.

\subsection{Joint Rician Posterior-Mean Benchmark}
\label{subsec:joint_rician_mmse}

The scalar Rician estimator in
\eqref{eq:scalar_rician_posterior_mean} conditions on
individual LS channel coefficients and does not exploit
the correlations induced by the pilot matrix.
We therefore consider the joint Rician posterior-mean
estimator, which retains the complete LS noise covariance.

For receive row $i$, define
\begin{equation}
    \bm r_i =
    \widehat{\bm H}_{\mathrm{LS},i:}^{H},
    \qquad
    \bm h_i = \bm H_{i:}^{H}.
\end{equation}
The corresponding observation model is
\begin{equation}
    \bm r_i = \bm h_i + \bm e_i,
    \qquad
    \bm e_i \sim
    \mathcal N_{\mathrm C}(\bm 0,\bm\Sigma),
\end{equation}
where
\begin{equation}
    \bm\Sigma =
    \sigma_n^2(\bm X\bm X^H)^{-1}.
    \label{eq:joint_ls_covariance}
\end{equation}
Unlike the scalar estimator, the joint estimator
retains the off-diagonal entries of $\bm\Sigma$.

Under the random-phase Rician channel model,
write
\begin{equation}
    \bm h_i = \bm m_i + \bm g_i,
\end{equation}
where
\begin{equation}
    \bm m_i =
    a\begin{bmatrix}
        e^{j\theta_1}\\
        e^{j\theta_2}
    \end{bmatrix},
    \qquad
    \bm g_i \sim
    \mathcal N_{\mathrm C}(\bm 0,v\bm I_2),
\end{equation}
with $a=\sqrt{K/(K+1)}$ and $v=1/(K+1)$.
The phases $\theta_1$ and $\theta_2$ are independent
and uniformly distributed over $[0,2\pi)$.

Define
\begin{equation}
    \bm A=(v\bm I_2+\bm\Sigma)^{-1},
    \qquad
    \bm b=\bm A\bm r_i.
\end{equation}
Conditioning on the Rician phases and integrating
the Gaussian component analytically gives~\cite{Kay1993}
\begin{equation}
\begin{split}
    \widehat{\bm h}_{i,\mathrm J}
    ={}& v\bm b\\
    &+(\bm I_2-v\bm A)
    \mathbb E[\bm m_i\mid\bm r_i].
\end{split}
    \label{eq:joint_rician_mmse}
\end{equation}

The remaining posterior expectation involves two
random phases. However, one phase can be integrated
analytically. For fixed $\theta=\theta_1$, define
\begin{equation}
    c(\theta)=
    ab_2-a^2 A_{21}e^{j\theta}.
\end{equation}
The conditional distribution of $\theta_2$ is
von Mises~\cite{Mardia1999}, with circular moment
\begin{equation}
\begin{split}
    \eta(\theta)
    &=
    \mathbb E[e^{j\theta_2}\mid\theta,\bm r_i]\\
    &=
    \frac{c(\theta)}{|c(\theta)|}
    \frac{I_1(2|c(\theta)|)}
         {I_0(2|c(\theta)|)},
\end{split}
\end{equation}
where $\eta(\theta)=0$ when $c(\theta)=0$.

The unnormalized marginal phase weight is
\begin{equation}
\begin{split}
    w(\theta)
    ={}&
    \exp\!\left(
    2a\operatorname{Re}
    \{b_1e^{-j\theta}\}
    \right)\\
    &\times I_0(2|c(\theta)|).
\end{split}
\end{equation}
Consequently,
\begin{equation}
    \mathbb E[\bm m_i\mid\bm r_i]
    =
    \frac{a}{Z}
    \int_0^{2\pi}
    w(\theta)
    \begin{bmatrix}
        e^{j\theta}\\
        \eta(\theta)
    \end{bmatrix}
    d\theta,
    \label{eq:joint_rician_phase_integral}
\end{equation}
where
\begin{equation}
    Z=\int_0^{2\pi}w(\theta)\,d\theta.
\end{equation}

Equation~\eqref{eq:joint_rician_phase_integral}
requires only one-dimensional periodic quadrature.
The Gaussian component and one phase are integrated
analytically, reducing the quadrature complexity
from $\mathcal O(Q^2)$ to $\mathcal O(Q)$ for
$Q$ integration nodes.

The estimator in \eqref{eq:joint_rician_mmse}
is the Bayesian posterior mean under the assumed
joint Rician channel and observation models.
For diagonal $\bm\Sigma$, it reduces to the
coefficientwise scalar Rician estimator.
At $K=0$, it reduces to the Gaussian LMMSE estimator
in \eqref{eq:lmmse_channel_estimate}.

We evaluate the remaining integral using a periodic
trapezoidal rule with numerically stabilized Bessel
functions and normalized exponential weights.
The numerical implementation uses $Q=256$ nodes,
with convergence verified against $Q=512$.

\subsection{Physical Latent Channel Parameterization}

The latent vector is defined directly from the Cartesian components of the
complex channel coefficients as
\begin{equation}
\begin{aligned}
\bm z
=
\sqrt{2}
\big[
&\Re\{h_{HH}\},\,
\Im\{h_{HH}\},\,
\Re\{h_{HV}\},\,
\Im\{h_{HV}\},\\
&\Re\{h_{VH}\},\,
\Im\{h_{VH}\},\,
\Re\{h_{VV}\},\,
\Im\{h_{VV}\}
\big]^{T}.
\end{aligned}
\label{eq:latent_channel_vector}
\end{equation}
The corresponding channel matrix is recovered through the fixed physical map
\begin{equation}
\bm H(\bm z)
=
\frac{1}{\sqrt{2}}
\begin{bmatrix}
z_1+jz_2 & z_3+jz_4\\
z_5+jz_6 & z_7+jz_8
\end{bmatrix}.
\label{eq:latent_to_channel}
\end{equation}
Hence, the latent variables have a direct physical interpretation and the
latent-to-channel mapping contains no learned decoder parameters. The learned
component solves the inverse inference problem from the observed channel
statistics to the approximate posterior over $\bm z$, whereas
$\bm z \mapsto \bm H$ is already specified by the physical channel
parameterization. We therefore retain this known physical mapping explicitly
rather than approximate it with a learned decoder. This construction removes
learned decoder flexibility as a confounding factor in the study of
latent prior mismatch.
\begin{figure}[b!]
\centering
\resizebox{\columnwidth}{!}{%
\begin{tikzpicture}[
    >=Latex,
    block/.style={
        draw,
        rounded corners,
        align=center,
        text width=3.3cm,
        minimum height=1.45cm,
        fill=gray!10,
        font=\small,
        inner sep=3pt
    },
    arrow/.style={->,thick}
]

\node[block] (stats) at (0,0)
{Channel statistics\\[0.5mm]
$\widehat{\bm H}_{\mathrm{LS}}$\\
$\bm S_{xx}/N_{\mathrm{est}}$};

\node[block] (features) at (4.3,0)
{Feature stacking\\
and standardization\\[0.5mm]
$\bm s$};

\node[block] (encoder) at (8.6,0)
{Variational encoder\\[0.5mm]
$\bm\mu_\phi(\bm s)$\\
$\log\bm\sigma_\phi^2(\bm s)$};

\node[block] (physical) at (0,-2.3)
{Fixed physical mapping\\[0.5mm]
$\widehat{\bm H}=\bm H(\bm z)$};

\node[block] (reparam) at (4.3,-2.3)
{Reparameterization\\[0.5mm]
$\bm z=\bm\mu_\phi+
\bm\sigma_\phi\odot\bm\epsilon$\\
$\bm\epsilon\sim\mathcal N(\bm0,\bm I)$};

\node[block] (posterior) at (8.6,-2.3)
{Approximate posterior\\[0.5mm]
$q_\phi(\bm z\mid\bm s)$};

\node[block] (predict) at (0,-4.6)
{Structural output\\[0.5mm]
$\widehat{\bm Y}=\widehat{\bm H}\bm X$};

\node[block] (rec) at (4.3,-4.6)
{Reconstruction term\\[0.5mm]
$\displaystyle
\begin{aligned}
\mathcal L_{\mathrm{rec}}
&=\frac{1}{\sigma_n^2}\\[-0.5mm]
&\hspace{-6mm}\operatorname{mean}
\left(|\widehat{\bm H}\bm X-\bm Y|^2\right)
\end{aligned}$};

\node[block] (objective) at (8.6,-4.6)
{Training objective\\[0.5mm]
$\displaystyle
\begin{aligned}
\mathcal L={}&\mathcal L_{\mathrm{rec}}\\
&+\zeta D_{\mathrm{KL}}(q_\phi\Vert p)
\end{aligned}$};

\node[block] (prior) at (4.3,-6.9)
{Latent prior $p(\bm z)$\\[0.5mm]
Gaussian, Rician,\\
Laplace, or Cauchy};

\node[block] (KL) at (8.6,-6.9)
{Forward KL regularization\\[0.5mm]
$\displaystyle
D_{\mathrm{KL}}\left(
q_\phi(\bm z\mid\bm s)
\Vert p(\bm z)
\right)$};

\draw[arrow] (stats.east) -- (features.west);
\draw[arrow] (features.east) -- (encoder.west);
\draw[arrow] (encoder.south) -- (posterior.north);
\draw[arrow] (posterior.west) -- (reparam.east);
\draw[arrow] (reparam.west) -- (physical.east);
\draw[arrow] (physical.south) -- (predict.north);
\draw[arrow] (predict.east) -- (rec.west);
\draw[arrow] (rec.east) -- (objective.west);

\draw[arrow] (prior.east) -- (KL.west);
\draw[arrow] (KL.north) -- (objective.south);

\draw[arrow]
(posterior.east) -- ++(0.65,0)
|- (KL.east);

\end{tikzpicture}}
\caption{Block diagram of the variational channel estimator. The encoder maps
the \ac{LS} channel estimate and normalized pilot Gram matrix to a
diagonal-Gaussian posterior over the physical latent channel coordinates.
Latent samples are converted to the estimated $2\times2$ complex channel
through the fixed mapping $\widehat{\bm H}=\bm H(\bm z)$.
Training combines the reconstruction term with the forward \ac{KL} penalty
weighted by $\zeta$. The assumed latent prior is varied among the
conventional isotropic Gaussian model, Rician priors, and the
Laplace and Cauchy control priors.}
\label{fig:prior_variational_estimator}
\end{figure}
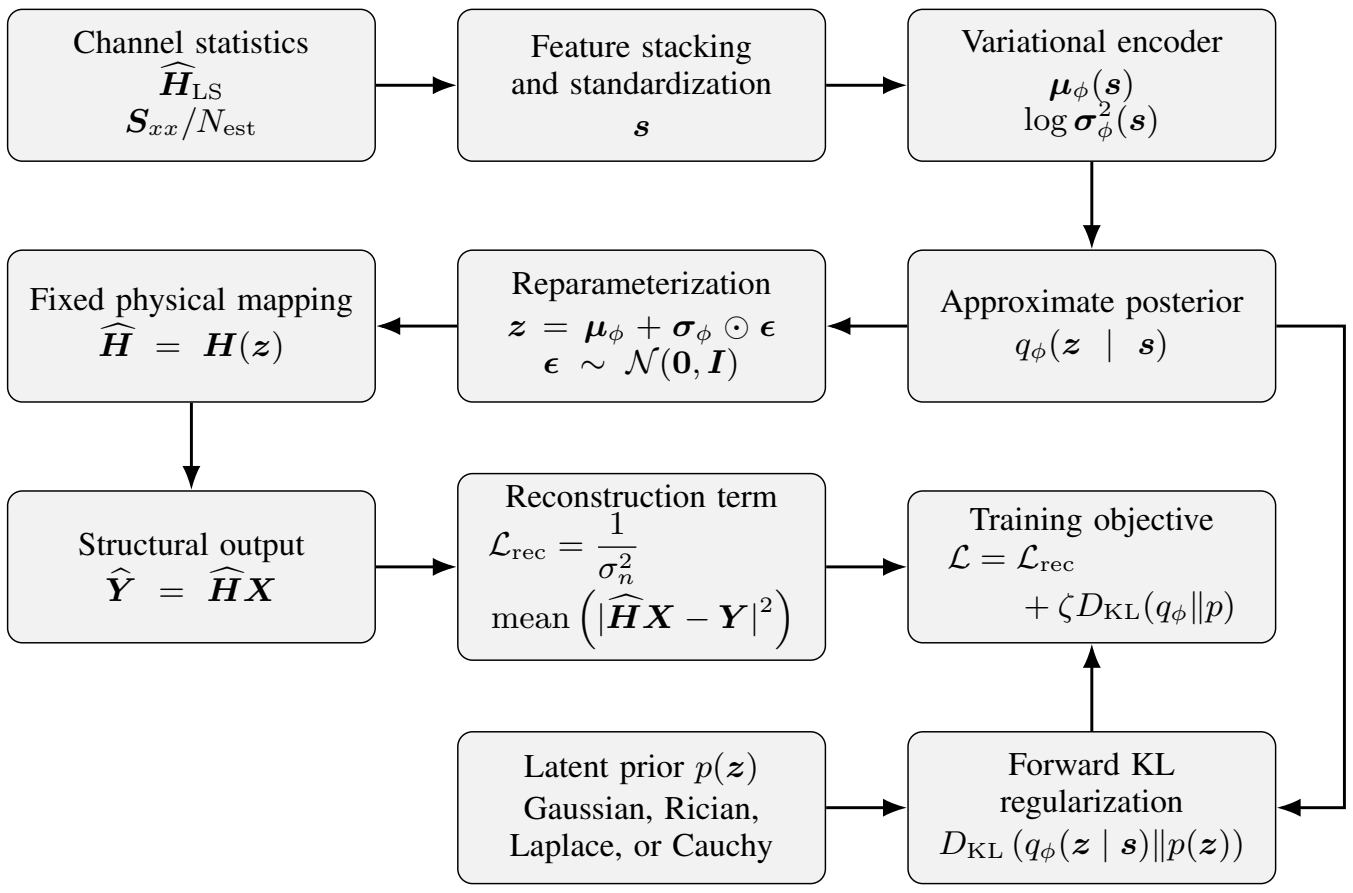

\vspace{-3mm}
\section{Variational Channel Estimator}
\label{sec:vae}

The proposed receiver uses a variational estimator to infer the physical
dual-polarization channel from channel-relevant second-order statistics.
Rather than learning a generative representation of the received signal, the
estimator represents the normalized Cartesian channel coefficients by a latent
random variable and infers the corresponding posterior distribution.
Fig.~\ref{fig:prior_variational_estimator} summarizes the estimator
architecture and the associated reconstruction and prior-regularization paths.

For each estimation block, the encoder input is formed from the \ac{LS} channel
estimate $\widehat{\bm H}_{\mathrm{LS}}$ and the normalized pilot Gram matrix
$\bm S_{xx}/N_{\mathrm{est}}$, as defined in Section~\ref{sec:model}.
Their real and imaginary components are stacked into a $16$-dimensional
real feature vector $\bm s$, which is standardized using training-set
statistics before being supplied to the variational encoder.

The encoder parameterizes a diagonal Gaussian approximate posterior
\begin{equation}
    q_{\phi}(\bm z \mid \bm s)
    =
    \mathcal{N}
    \left(
        \bm \mu_{\phi}(\bm s),
        \operatorname{diag}
        \left(
            \bm \sigma_{\phi}^{2}(\bm s)
        \right)
    \right),
    \label{eq:vae_posterior}
\end{equation}
where $\phi$ denotes the encoder parameters. Latent samples are obtained
through the reparameterization
\begin{equation}
    \bm z
    =
    \bm \mu_{\phi}
    +
    \bm \sigma_{\phi}
    \odot
    \bm \epsilon,
    \qquad
    \bm \epsilon\sim\mathcal{N}(\bm 0,\bm I).
    \label{eq:reparameterization}
\end{equation}

Although the learned generative decoder of a conventional \ac{VAE} is absent, the
estimator retains amortized variational inference, a parameterized approximate
posterior, reparameterized latent sampling, and \ac{KL} regularization. The known
physical map $\bm z\mapsto\bm H$ replaces the learned decoder. The sampled
latent vector is therefore converted directly to the complex channel matrix
through the fixed physical mapping in \eqref{eq:latent_to_channel},
\begin{equation}
    \widehat{\bm H}
    =
    \bm H(\bm z).
    \label{eq:vae_channel_mapping}
\end{equation}
The estimated channel is then inserted into the physical observation model,
\begin{equation}
    \widehat{\bm Y}
    =
    \widehat{\bm H}\bm X.
    \label{eq:vae_prediction}
\end{equation}
Hence, the latent variables retain a direct interpretation as normalized
channel coefficients throughout the estimation procedure.

The reconstruction term is evaluated in the observation domain as
\begin{equation}
    \mathcal{L}_{\mathrm{rec}}
    =
    \frac{1}{\sigma_n^2}
    \operatorname{mean}
    \left(
        \left|
            \widehat{\bm H}\bm X-\bm Y
        \right|^2
    \right),
    \label{eq:vae_reconstruction}
\end{equation}
where $\sigma_n^2$ denotes the complex receiver-noise variance.
The mean in \eqref{eq:vae_reconstruction} is taken over the
$2N_{\mathrm{est}}$ complex residual entries per block and
over the minibatch. The normalization expresses the reconstruction
error relative to the nominal noise level. The forward \ac{KL} divergence
is summed over the eight latent dimensions for each block and then
averaged over the minibatch. Because $\widehat{\bm H}$ is obtained
through the fixed map $\bm H(\bm z)$, the reconstruction term
directly penalizes errors in a physically interpretable channel
representation.

The estimator is regularized using the forward Kullback--Leibler divergence
between the approximate posterior and the assumed latent prior,
\begin{equation}
    D_{\mathrm{KL}}
    \left(
        q_{\phi}(\bm z\mid\bm s)
        \Vert
        p(\bm z)
    \right).
    \label{eq:forward_KL}
\end{equation}
The complete training objective is
\begin{equation}
    \mathcal{L}
    =
    \mathcal{L}_{\mathrm{rec}}
    +
    \zeta
    D_{\mathrm{KL}}
    \left(
        q_{\phi}(\bm z\mid\bm s)
        \Vert
        p(\bm z)
    \right),
    \label{eq:vae_objective}
\end{equation}
where $\zeta\geq0$ controls the influence of the latent prior on the
estimated posterior.

For the proper complex Gaussian observation model in
\eqref{eq:block_model}, the reconstruction term in
\eqref{eq:vae_reconstruction} averages the normalized
squared residuals over $2N_{\mathrm{est}}$ complex observations. Consequently, choosing $\zeta=1/(2N_{\mathrm{est}})$ makes ~\eqref{eq:vae_objective} equivalent to the negative evidence lower bound (ELBO), scaled by $1/(2N_{\mathrm{est}})$ and up to terms independent of the variational parameters. Thus, the regularization weight follows from the observation likelihood rather than hyperparameter tuning.

The form of $p(\bm z)$ is the central variable in the present study. The
conventional estimator uses the isotropic Gaussian prior
$p_{\mathrm G}(\bm z)=\mathcal{N}(\bm 0,\bm I)$, while the physically matched
estimator uses the Rician prior associated with the channel law. We additionally
vary the assumed Rician factor to study prior-parameter mismatch and consider
Laplace and Cauchy priors as generic non-Gaussian controls. The posterior
family, estimator architecture, observation model, latent-to-channel mapping,
and forward \ac{KL} objective are otherwise kept unchanged. This comparison measures the effect of
prior choice within the specified amortized variational estimator.

\section{Physical Latent Prior Models}
\label{sec:prior}

The variational estimator in Section~\ref{sec:vae} uses the forward
Kullback--Leibler divergence throughout. We distinguish between the physical
latent distribution $p_{\mathrm{physical}}(\bm z)$, the prior used in the
variational regularizer $p_{\mathrm{prior}}(\bm z)$, and the approximate
posterior $q_{\phi}(\bm z\mid\bm s)$. In the matched case,
$p_{\mathrm{prior}}=p_{\mathrm{physical}}$, whereas in the mismatch and
assumed-$K$ experiments these distributions differ. Hence, the form of the
training objective is fixed and only the assumed latent prior is varied. This
allows the effect of prior mismatch to be studied without changing the
divergence, estimator architecture, or observation model. For brevity, we
write $p(\bm z)$ for the assumed latent prior $p_{\mathrm{prior}}(\bm z)$ in
the estimator equations and figures.

We use $K_{\mathrm{true}}$ for the Rician factor that generates the physical
channel distribution and $K_{\mathrm{prior}}$ for the Rician factor assumed
by the latent prior in the \ac{KL} term.

\subsection{Isotropic Gaussian Prior}

The conventional latent prior is
\begin{equation}
    p_{\mathrm{G}}(\bm z)
    =
    \mathcal{N}
    \left(
        \bm 0,
        \bm I_8
    \right).
    \label{eq:gaussian_prior}
\end{equation}
Under the Cartesian parameterization in
\eqref{eq:latent_channel_vector}, this prior corresponds to independent
zero-mean Gaussian real and imaginary channel components with equal variance.
Equivalently, the complex channel coefficients are independent proper
complex Gaussian random variables with unit average power.

This model provides the matched reference at the Rayleigh limit. It also
provides the mismatched reference when the physical channel contains a
nonzero Rician line-of-sight component.

\subsection{Rician Prior}

To introduce a controlled departure from the Gaussian channel law, each
complex channel coefficient is generated according to
\begin{equation}
    h
    =
    \sqrt{\frac{K}{K+1}}\,e^{j\theta}
    +
    \sqrt{\frac{1}{K+1}}\,g,
    \label{eq:rician_channel}
\end{equation}
where
\begin{equation}
    g\sim\mathcal{N}_{\mathrm{C}}(0,1),
    \qquad
    \theta\sim\mathcal{U}[0,2\pi).
\end{equation}
The random-phase Rician model is also considered in
\cite{Ozdogan2019} in the context of channel estimation
for cell-free massive MIMO systems. 
The normalization in \eqref{eq:rician_channel} gives
$\mathbb{E}[|h|^2]=1$ for every value of $K$. The Rician factor therefore
changes the relative strength of the line-of-sight and diffuse components
without changing the average coefficient power.

For one complex coefficient, define the corresponding two-dimensional
Cartesian latent vector as
\begin{equation}
    \bm u
    =
    \sqrt{2}
    \begin{bmatrix}
        \Re\{h\} &
        \Im\{h\}
    \end{bmatrix}^{T},
    \qquad
    \rho=\|\bm u\|_2.
\end{equation}
After marginalizing over the uniformly distributed phase $\theta$, the
log-density of $\bm u$ is
\begin{equation}
\begin{split}
    \log p_K(\bm u)
    ={}&
    \log(K+1)
    -\log(2\pi)
    -K \\
    &-\frac{K+1}{2}\rho^2
    +\log I_0
    \left(
        \sqrt{2K(K+1)}\,\rho
    \right),
\end{split}
\label{eq:rician_latent_density}
\end{equation}
where $I_0(\cdot)$ denotes the modified Bessel function of the first kind.
The complete eight-dimensional prior is obtained by applying
\eqref{eq:rician_latent_density} independently to the four Cartesian
coefficient pairs,
\begin{equation}
    \log p_K(\bm z)
    =
    \sum_{m=1}^{4}
    \log p_K(\bm u_m).
    \label{eq:rician_full_prior}
\end{equation}

At $K=0$, \eqref{eq:rician_latent_density} reduces to
\begin{equation}
    \log p_0(\bm u)
    =
    -\log(2\pi)
    -\frac{1}{2}\|\bm u\|_2^2,
\end{equation}
and hence
\begin{equation}
    p_0(\bm z)
    =
    \mathcal{N}
    \left(
        \bm0,
        \bm I_8
    \right).
    \label{eq:rician_rayleigh_limit}
\end{equation}
Thus, the Rician prior and the conventional Gaussian prior are identical at
the Rayleigh limit. Increasing $K$ introduces a progressively stronger
departure from the Gaussian reference while preserving the same latent
coordinates and average channel power.

In the matched case, the Rician factor used in the assumed prior equals the
factor used to generate the physical channel,
$K_{\mathrm{prior}}=K_{\mathrm{true}}$. To study prior-parameter mismatch,
we also evaluate Rician priors with
$K_{\mathrm{prior}}\neq K_{\mathrm{true}}$ while keeping the physical channel
law and all other estimator components fixed.

\subsection{Non-Gaussian Control Priors}

For the control experiment, the factorized Laplace prior is
\begin{equation}
    p_{\mathrm{L}}(\bm z)
    =
    \prod_{i=1}^{8}
    \frac{1}{2b}
    \exp\!\left(-\frac{|z_i|}{b}\right),
    \qquad
    b=\frac{1}{\sqrt{2}}.
    \label{eq:laplace_prior}
\end{equation}
Thus, $\operatorname{Var}(z_i)=2b^2=1$, so the Laplace control has the
same marginal coordinate variance as the standard Gaussian latent prior.
The factorized Cauchy control is
\begin{equation}
    p_{\mathrm{C}}(\bm z)
    =
    \prod_{i=1}^{8}
    \frac{1}{\pi\gamma\left[1+(z_i/\gamma)^2\right]},
    \qquad
    \gamma=\frac{1}{\sqrt{2}}.
    \label{eq:cauchy_prior}
\end{equation}
The Cauchy prior has no finite variance and is therefore not
variance-matched to the Gaussian or Laplace priors. It is used only as a
generic infinite-variance heavy-tail stress-test prior. Neither control prior
is claimed to represent the physical Rician channel law.

\subsection{Forward KL Evaluation}

For the Gaussian prior, the \ac{KL} term between the diagonal-Gaussian posterior
and $p_{\mathrm{G}}(\bm z)$ can be evaluated in closed form. For a Rician prior, we evaluate the same forward \ac{KL} objective as
\begin{equation}
    D_{\mathrm{KL}}
    \left(
        q_{\phi}\Vert p_K
    \right)
    =
    \mathbb{E}_{q_{\phi}}
    \left[
        \log q_{\phi}(\bm z\mid\bm s)
        -
        \log p_K(\bm z)
    \right].
    \label{eq:rician_forward_KL}
\end{equation}
The expectation is approximated using latent samples generated through
\eqref{eq:reparameterization}. The Rician log-density is evaluated using a
numerically stable implementation of $\log I_0(\cdot)$.
The Laplace and Cauchy control priors use the same sampled forward \ac{KL} path,
with only the evaluated log-prior density replaced by
$\log p_{\mathrm{L}}(\bm z)$ or $\log p_{\mathrm{C}}(\bm z)$,
respectively.

Across all prior comparisons, the estimator components other than the assumed
latent prior are kept fixed. The Gaussian estimator uses
$p_{\mathrm{G}}(\bm z)$, the Rician experiments use $p_K(\bm z)$ with either
matched or deliberately mismatched $K_{\mathrm{prior}}$, and the control
experiments use $p_{\mathrm{L}}(\bm z)$ or $p_{\mathrm{C}}(\bm z)$. The
approximate posterior, forward \ac{KL} direction, reconstruction term, physical
latent-to-channel mapping, observation model, and estimator architecture
remain unchanged.

The regularization weight $\zeta$ controls the relative
contribution of the assumed prior to the variational objective.
For small $\zeta$, the reconstruction term dominates, whereas
increasing $\zeta$ gives greater weight to prior regularization.
In the main experiments, we use the physically calibrated
weight $\zeta=1/(2N_{\mathrm{est}})$, corresponding to the
Gaussian observation model. Section~\ref{sec:results}
examines prior mismatch across Rician $K$-factors and
estimation-block lengths, sensitivity to the assumed Rician
factor, and generic non-Gaussian control priors.

Generalized variational inference distinguishes the loss function,
divergence measure, and variational family \cite{Knoblauch2022}.
In our case, the assumed prior also acts as a regularizer through
the \ac{KL} term. Consequently, the prior that minimizes channel
estimation error need not coincide with the physical channel
distribution.

\section{Training and Estimation Procedure}
\label{sec:training}

The variational estimator is trained from blockwise dual-polarization
observations using the channel-relevant statistics defined in
Section~\ref{sec:model}. We first compute the \ac{LS} channel estimate
$\widehat{\bm H}_{\mathrm{LS}}$ and normalize the pilot Gram matrix
$\bm S_{xx}$ by $N_{\mathrm{est}}$. Their real and imaginary components
are then stacked into a $16$-dimensional real feature vector and
standardized using training-set statistics. The same transformation
is applied at evaluation time.

\subsection{Training Procedure}

For each standardized feature vector $\bm s$, the encoder parameterizes the
diagonal-Gaussian posterior in \eqref{eq:vae_posterior}. Latent samples are
generated using \eqref{eq:reparameterization} and converted to
$\widehat{\bm H}$ through the fixed physical mapping in
\eqref{eq:vae_channel_mapping}. The resulting channel estimate is inserted
into the observation model in \eqref{eq:vae_prediction}.

Training minimizes the forward \ac{KL} objective in \eqref{eq:vae_objective}. The
regularization weight $\zeta$ controls the influence of the assumed latent
prior independently of the estimator architecture. The encoder parameters are
optimized using the reconstruction and \ac{KL} terms, while the latent-to-channel
mapping remains fixed. Numerical training settings are given in
Section~\ref{sec:simulations}.

\subsection{Channel Estimate and Evaluation}

At evaluation time, the posterior mean is mapped directly to the physical
channel estimate,
\begin{equation}
    \widehat{\bm H}
    =
    \bm H\!\left(
        \bm\mu_{\phi}(\bm s)
    \right),
    \label{eq:posterior_mean_channel}
\end{equation}
which removes latent-sampling variation from the reported estimate.
Channel estimation accuracy is measured by
\begin{equation}
    \operatorname{NRMSE}
    =
    \frac{
        \left\|\widehat{\bm H}-\bm H\right\|_{F}
    }{
        \left\|\bm H\right\|_{F}+\epsilon_{\mathrm{N}}
    },
    \label{eq:channel_nrmse}
\end{equation}
where $\epsilon_{\mathrm{N}}>0$ is a small numerical constant. For $B$ test blocks, the \ac{MSFE} is
\begin{equation}
    \mathrm{MSFE}
    =
    \frac{1}{B}
    \sum_{b=1}^{B}
    \left\|
        \widehat{\bm H}_b-\bm H_b
    \right\|_F^2,
    \label{eq:channel_msfe}
\end{equation}
where $\widehat{\bm H}_b$ and $\bm H_b$ denote the estimated
and true channel matrices for block $b$, respectively.

\section{Simulation Benchmark}
\label{sec:simulations}

\subsection{Simulation Configuration}

The benchmark uses the dual-polarization model in
\eqref{eq:system_model}. Each realization contains four independently
generated complex channel coefficients, remains constant over one estimation
block, and changes independently between blocks. The transmitted estimation
symbols satisfy
$\bm x_t\sim\mathcal{N}_{\mathrm C}(\bm0,\bm I_2)$, and the receiver noise is
proper complex Gaussian with $\sigma_n^2=0.02$. 

The transmitted estimation symbols represent known pilot
observations used for receiver channel estimation. The Gaussian
pilot distribution is a simulation assumption and does not
represent the quantum signal states used for key distribution.
We evaluate the effective channel estimation problem rather
than a complete \ac{CVQKD} transmission protocol.

\subsection{Rician Prior Mismatch Regime Study}

We use $N_{\mathrm{est}}\in\{32,48,64,128\}$,
$K\in\{10,30,100\}$, and the likelihood-derived regularization weight
$\zeta=1/(2N_{\mathrm{est}})$, giving $12$ operating points.
For each of three paired training seeds,
we generate $100{,}000$ independent training blocks and train each prior
condition for $300{,}000$ Adam updates with batch size $512$.

For all four pilot lengths, the encoder uses two fully connected
hidden layers of width $424$ with ReLU activations, followed by a
linear layer producing eight posterior means and eight
log-variances. Each predicted log-variance is clipped as
$\ell_d=\min\{20,\max\{-20,\widetilde{\ell}_d\}\}$, and the diagonal
posterior variance is $\sigma_d^2=\exp(\ell_d)$.
Each Adam update uses a minibatch of $512$ blocks and one
reparameterized latent sample per minibatch entry,
$\bm z=\bm\mu+\exp(\bm\ell/2)\odot\bm\epsilon$,
where $\bm\epsilon\sim\mathcal N(\bm0,\bm I_8)$.
The same sample evaluates the reconstruction term and the
matched Rician forward \ac{KL}. The Gaussian forward \ac{KL} is evaluated analytically. For update index $t$, the learning rate is piecewise constant
\begingroup
\renewcommand{\arraystretch}{0.9}
\begin{equation}
\eta_t=
\left\{
\begin{array}{@{}l@{\quad}l@{}}
2\times10^{-3}, & 1\le t\le 3000,\\
2\times10^{-4}, & 3000<t\le 10000,\\
10^{-4}, & 10000<t\le 20000,\\
5\times10^{-5}, & 20000<t\le 30000,\\
2\times10^{-5}, & 30000<t\le 100000,\\
10^{-5}, & 100000<t\le 200000,\\
5\times10^{-6}, & 200000<t\le 300000.
\end{array}
\right.
\end{equation}
\endgroup

Checkpoints are selected separately for the Gaussian and matched Rician
models by minimum \ac{MSFE} on $5,000$ validation blocks. The selected models are
then evaluated on $10{,}000$ independent held-out test blocks. Within each
operating point, the physical test blocks are shared across priors and
training seeds; each seed uses an independent training population with
paired initialization, minibatch, and latent-sampling streams across priors.
The matched prior uses the physical Rician factor, whereas the Gaussian
condition uses $p_{\mathrm G}(\bm z)$ in \eqref{eq:gaussian_prior}.
The $K=0$ reference is analytical prior equality only. No neural experiment is performed at $K=0$.

We evaluate the scalar and joint Rician posterior-mean
estimators on the same $10{,}000$ held-out blocks at each
of the $12$ operating points. The joint estimator uses
the complete complex \ac{LS} noise covariance and
one-dimensional periodic quadrature with $Q=256$ nodes.
We verify numerical convergence by repeating the
posterior calculations with $Q=512$ nodes. All $120{,}000$
held-out blocks satisfy the convergence criterion, with
a maximum posterior discrepancy of $9.21\times10^{-15}$
in Frobenius norm. The maximum corresponding correction
in per-block squared estimation error is
$7.88\times10^{-16}$. The analytical estimators use
the original physical test observations without
additional simulations or neural retraining.

\subsection{Assumed Rician Factor Sensitivity}

To examine prior parameter mismatch, the physical channel is generated with
$K_{\mathrm{true}}\in\{30{,}100\}$ while the assumed prior uses
\[
\frac{K_{\mathrm{prior}}}{K_{\mathrm{true}}}
\in\{1,2,4,8,16\},
\]
for $N_{\mathrm{est}}\in\{64{,}128\}$ and $\zeta=1/(2N_{\mathrm{est}})$. The ratio one gives
the physically matched prior. Changing $K_{\mathrm{prior}}$ affects only the
assumed latent prior.
\begin{figure*}[t!]
    \centering
    \subfloat[Median paired channel \ac{NRMSE} difference\label{fig:rician_delta_nrmse}]
    {\includegraphics[width=0.49\textwidth]{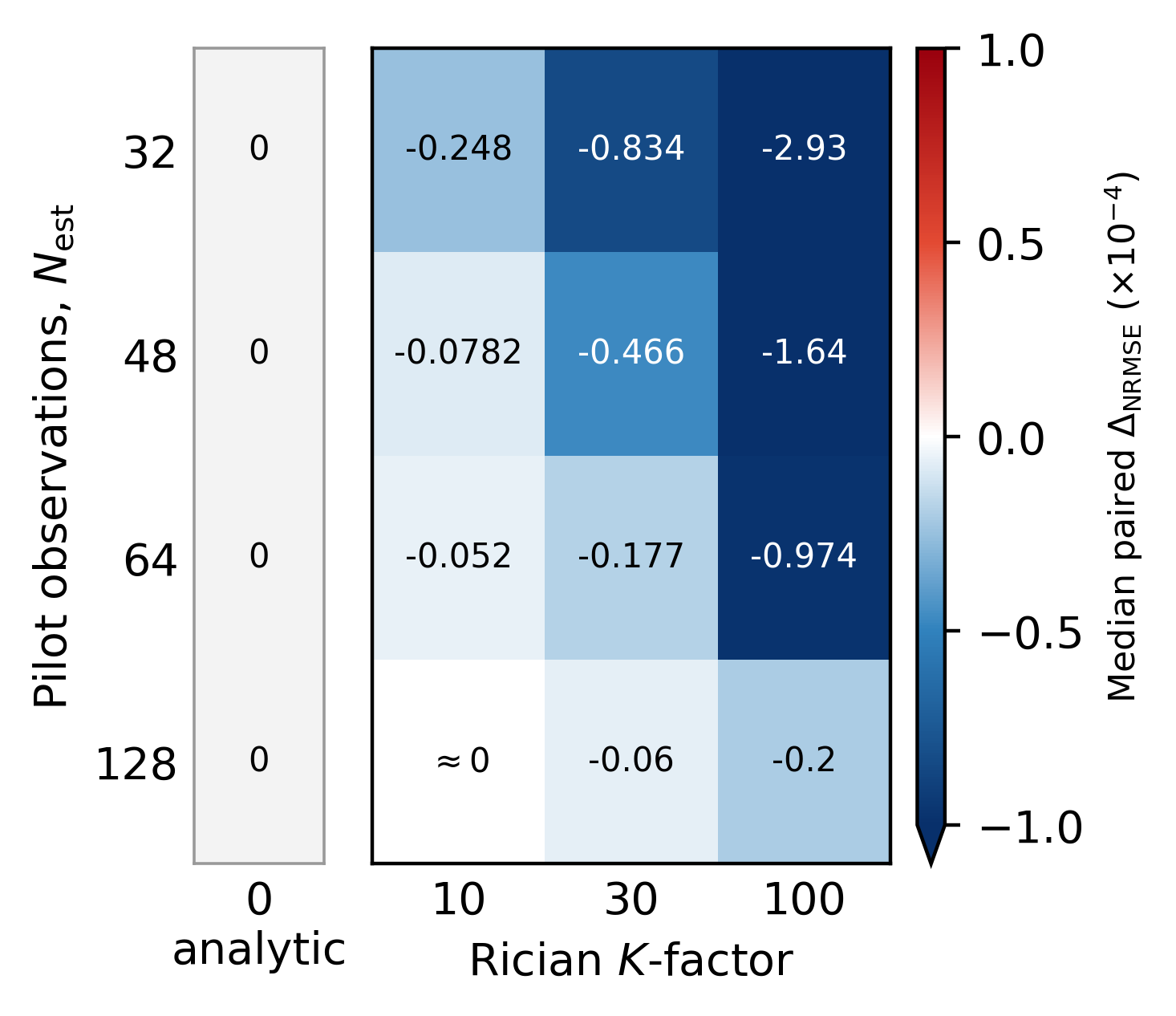}}
    \hfil
    \subfloat[Fraction of test blocks favoring the matched prior\label{fig:rician_win_fraction}]
    {\includegraphics[width=0.49\textwidth]{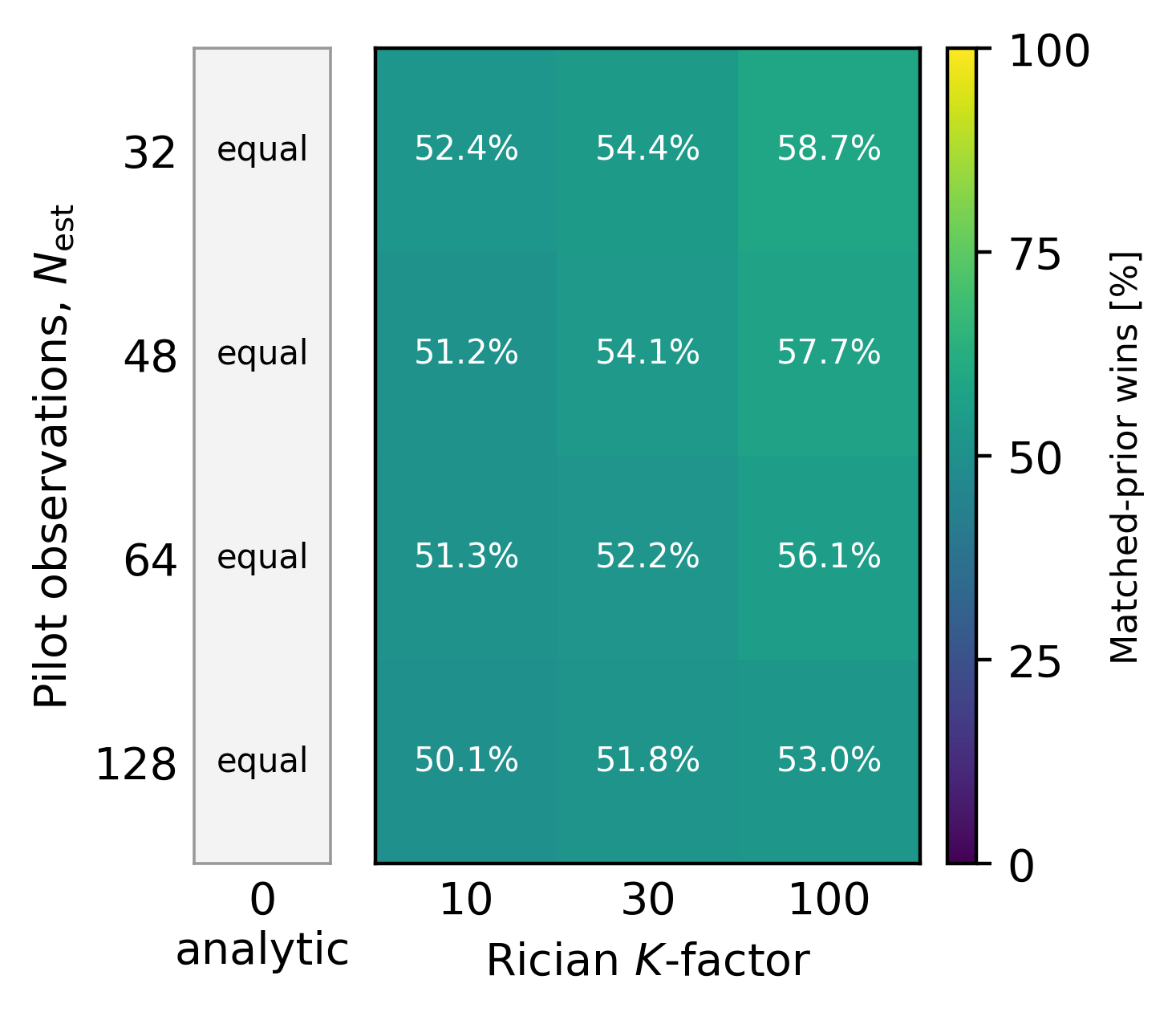}}
    \caption{Paired neural channel \ac{NRMSE} comparisons at
$\zeta=1/(2N_{\mathrm{est}})$. Both models are trained for $300{,}000$ updates
for each of three paired training seeds, with checkpoints selected
independently by \ac{MSFE} on $5,000$ validation blocks. For each seed, paired
statistics use the same $10{,}000$ held-out test blocks. Panel (a) shows the
median across the three seed-specific block medians of
$\Delta_{\mathrm{NRMSE}}=\mathrm{NRMSE}_{\mathrm{matched}}
-\mathrm{NRMSE}_{\mathrm{Gaussian}}$; negative values favor the matched prior.
Panel (b) shows the mean of the three seed-specific fractions with
$\Delta_{\mathrm{NRMSE}}<0$. Test blocks are shared across seeds; these
block statistics do not constitute independent training trials.
The boxed $K=0$ column is an analytical Gaussian-null reference: zero
difference and identical priors, with no empirical win fraction; it is
excluded from heatmap normalization. In panel (a), colors are centered at
zero and saturate outside $\pm10^{-4}$, as indicated by the colorbar arrow;
annotations retain the larger measured values. Cell medians with absolute
value below $10^{-6}$ are displayed as $\approx0$ only for readability.}
    \vspace{1mm}
    \label{fig:rician_regime_map}
\end{figure*}

\subsection{Non-Gaussian Prior Controls}

For $K_{\mathrm{true}}\in\{30{,}100\}$ and
$N_{\mathrm{est}}\in\{64,128\}$, we compare the Gaussian, Laplace, Cauchy,
physically matched Rician, and sharpened Rician priors. The sharpened Rician
conditions use $K_{\mathrm{prior}}/K_{\mathrm{true}}=8$ for
$K_{\mathrm{true}}=30$ and $4$ for $K_{\mathrm{true}}=100$.
These ratios provide additional tests of prior sharpening relative
to the physically matched Rician prior. All remaining settings
are unchanged.

\subsection{Receiver-Level Metrics and Monte Carlo Protocol}

The least-squares estimator in \eqref{eq:ls_channel_estimate} is included as a
nonvariational reference. Channel \ac{NRMSE} in \eqref{eq:channel_nrmse}
and \ac{MSFE} are used to evaluate channel estimation accuracy. For each training seed, pooled channel \ac{NRMSE} is computed as
\begin{equation}
    \mathrm{NRMSE}_{\mathrm{pooled}}
    =
    \sqrt{
        \frac{
            \sum_{b=1}^{B}
            \|\widehat{\bm H}_b-\bm H_b\|_F^2
        }{
            \sum_{b=1}^{B}
            \|\bm H_b\|_F^2
        }
    },
    \label{eq:pooled_nrmse}
\end{equation}
where $B=10{,}000$ denotes the number of held-out test blocks.
Fig.~\ref{fig:rician_k_prior_upper_sweep} reports the mean of this
quantity across three training seeds, with $95\%$ confidence
intervals based on Student's $t$ distribution with two degrees
of freedom. The squared errors and channel energies are pooled
within each seed, not across seeds.

Within each training seed, all compared prior conditions
use the same physical channel, transmitted signal, receiver noise, network
initialization, and stochastic latent sampling streams where applicable.

Results are summarized primarily by the median and interquartile range.
For the main matched-versus-Gaussian study,
\begin{equation}
    \Delta_{\mathrm{NRMSE}}
    =
    \operatorname{NRMSE}_{\mathrm{matched}}
    -
    \operatorname{NRMSE}_{\mathrm{Gaussian}},
    \label{eq:paired_delta_nrmse}
\end{equation}
so $\Delta_{\mathrm{NRMSE}}<0$ favors the physically matched Rician prior.

Other paired comparisons are defined analogously. Test-block quartiles
and IQRs are computed separately for each seed, while ranges and
standard deviations across the three seed-specific medians and win
fractions describe training-seed variation. The training seeds are
$20260817$, $20260819$, and $20260820$. The shared test blocks are
not pooled as independent training replications.

For comparisons involving the joint Rician \ac{MMSE}
estimator, we compute pointwise $95\%$ paired
block-bootstrap confidence intervals using $2{,}000$
resamples. For each neural estimator, squared estimation
errors are first averaged across the three training seeds
separately for each of the $10{,}000$ held-out blocks.
Each bootstrap replicate resamples $10{,}000$ block indices
with replacement, using identical indices across the
compared estimators. The paired \ac{MSFE} differences
are computed from the resampled losses, and the
$2.5$th and $97.5$th percentiles define the confidence
intervals. These intervals quantify test-block uncertainty
conditional on the three fixed trained models.
Training-seed uncertainty is reported separately using
Student's $t$ intervals with two degrees of freedom.

\section{Numerical Experiments}
\label{sec:results}

The numerical experiments evaluate the variational channel estimator under the
benchmark in Section~\ref{sec:simulations}. We first identify the analytical Gaussian-null reference and then examine the effects of Rician prior mismatch, channel estimation block length, assumed Rician factor, and generic
non-Gaussian control priors. We also compare the variational estimators
with analytical \ac{LS}, \ac{LMMSE}, scalar Rician posterior-mean,
and joint Rician \ac{MMSE} estimators. Channel \ac{NRMSE} and
\ac{MSFE} are used to evaluate estimation accuracy,
including paired comparisons between estimators.

\subsection{Gaussian Null Reference}
\label{subsec:gaussian_baseline}

At $K=0$, the Rician prior reduces exactly to
$p_{\mathrm G}(\bm z)=\mathcal{N}(\bm0,\bm I_8)$, as shown in
\eqref{eq:rician_rayleigh_limit}. Hence, no prior-law mismatch
is present. We use this analytical equality as the null reference
in Fig.~\ref{fig:rician_regime_map}. No separate neural
experiment is performed at $K=0$.

\begin{table*}[t!]
\centering
\caption{Absolute channel estimation \ac{MSFE} for \ac{LS}, \ac{LMMSE},
scalar matched Rician posterior mean, joint Rician \ac{MMSE},
Gaussian \ac{VAE}, and matched Rician \ac{VAE} at
$K_{\mathrm{true}}\in\{10,30,100\}$.
All \ac{MSFE} values are in units of $10^{-4}$.}
\label{tab:absolute_msfe_benchmarks}
\small
\setlength{\tabcolsep}{6pt}
\begin{tabular}{ccrrrrrrl}
\hline
$N_{\mathrm{est}}$ & $K_{\mathrm{true}}$
& LS & LMMSE & Scalar Rician & Joint Rician
& Gaussian VAE & Rician VAE
& $\Delta$ [95\% CI] \\
\hline
32  & 10  & 26.5011 & 26.4859 & 26.4116 & 26.4060 & 26.5631 & 26.5091
    & $+0.0232\;[+0.0086,+0.0378]$ \\
32  & 30  & 26.5976 & 26.5762 & 26.3413 & 26.3342 & 26.6423 & 26.4516
    & $-0.1246\;[-0.1646,-0.0846]$ \\
32  & 100 & 26.8239 & 26.8022 & 25.9354 & 25.9084 & 26.8716 & 26.1332
    & $-0.6690\;[-0.7439,-0.5941]$ \\
48  & 10  & 17.4656 & 17.4536 & 17.4266 & 17.4261 & 17.4972 & 17.4811
    & $+0.0275\;[+0.0062,+0.0488]$ \\
48  & 30  & 17.3866 & 17.3812 & 17.2618 & 17.2598 & 17.4220 & 17.3399
    & $-0.0413\;[-0.0734,-0.0091]$ \\
48  & 100 & 17.4690 & 17.4621 & 17.0732 & 17.0633 & 17.5090 & 17.1901
    & $-0.2720\;[-0.3074,-0.2367]$ \\
64  & 10  & 12.8809 & 12.8758 & 12.8555 & 12.8547 & 12.9140 & 12.9056
    & $+0.0298\;[+0.0031,+0.0564]$ \\
64  & 30  & 12.8428 & 12.8379 & 12.7780 & 12.7755 & 12.8801 & 12.8442
    & $+0.0063\;[-0.0059,+0.0185]$ \\
64  & 100 & 12.8358 & 12.8305 & 12.6303 & 12.6269 & 12.8642 & 12.7136
    & $-0.1169\;[-0.1295,-0.1044]$ \\
128 & 10  &  6.3563 &  6.3568 &  6.3550 &  6.3550 &  6.3831 &  6.3833
    & $+0.0265\;[+0.0074,+0.0457]$ \\
128 & 30  &  6.3430 &  6.3413 &  6.3278 &  6.3276 &  6.3710 &  6.3647
    & $+0.0234\;[+0.0022,+0.0445]$ \\
128 & 100 &  6.3620 &  6.3605 &  6.3165 &  6.3159 &  6.3872 &  6.3627
    & $+0.0022\;[-0.0130,+0.0174]$ \\
\hline
\end{tabular}
\end{table*}

\subsection{Rician Prior Mismatch Regime Study}
\label{subsec:rician_regime}

We compare matched Rician and isotropic Gaussian neural priors using
models trained for $300{,}000$ updates at
$N_{\mathrm{est}}\in\{32,48,64,128\}$,
$K\in\{10,30,100\}$, and $\zeta=1/(2N_{\mathrm{est}})$. For each seed, we first compute paired \ac{NRMSE} differences over the shared
held-out blocks. Fig.~\ref{fig:rician_delta_nrmse} reports the median of
these three seed-specific medians, and
Fig.~\ref{fig:rician_win_fraction} reports the mean of the three
seed-specific fractions of blocks with a strictly negative difference.
Both statistics use posterior-mean channel estimates selected by independent
validation \ac{MSFE}. The reported medians are computed directly from paired
per-block \ac{NRMSE} differences.

At $K=100$, the median paired \ac{NRMSE} differences are approximately
$-2.93\times10^{-4}$, $-1.64\times10^{-4}$,
$-9.74\times10^{-5}$, and $-2.00\times10^{-5}$ for
$N_{\mathrm{est}}=32$, $48$, $64$, and $128$, respectively.
The corresponding mean fractions of blocks favoring the matched prior are
$58.71\%$, $57.71\%$, $56.11\%$, and $52.97\%$.
At $K=10$ and $N_{\mathrm{est}}=128$, the displayed median difference is
approximately $+7.51\times10^{-8}$, and the seed-specific medians span
$-1.07\times10^{-6}$ to $+1.77\times10^{-7}$. This operating point is
close to a tie rather than a consistent matched-prior advantage.

Test-block variation and training-seed variation quantify different effects.
For example, at $K=100$ and $N_{\mathrm{est}}=32$, the paired-difference
IQR within each seed is between $1.79\times10^{-3}$ and
$1.84\times10^{-3}$, whereas the three block medians range only from
$-2.94\times10^{-4}$ to $-2.91\times10^{-4}$. Thus, a negative central
shift does not imply improvement on every test block. On this calibrated finite grid, the matched-prior shift is larger at
stronger LOS and with fewer pilot observations. The analytical $K=0$ column denotes equality of the prior
densities and is not an additional trained or Monte Carlo operating point.
\begin{figure}[b!]
\centering
\includegraphics[width=\columnwidth]{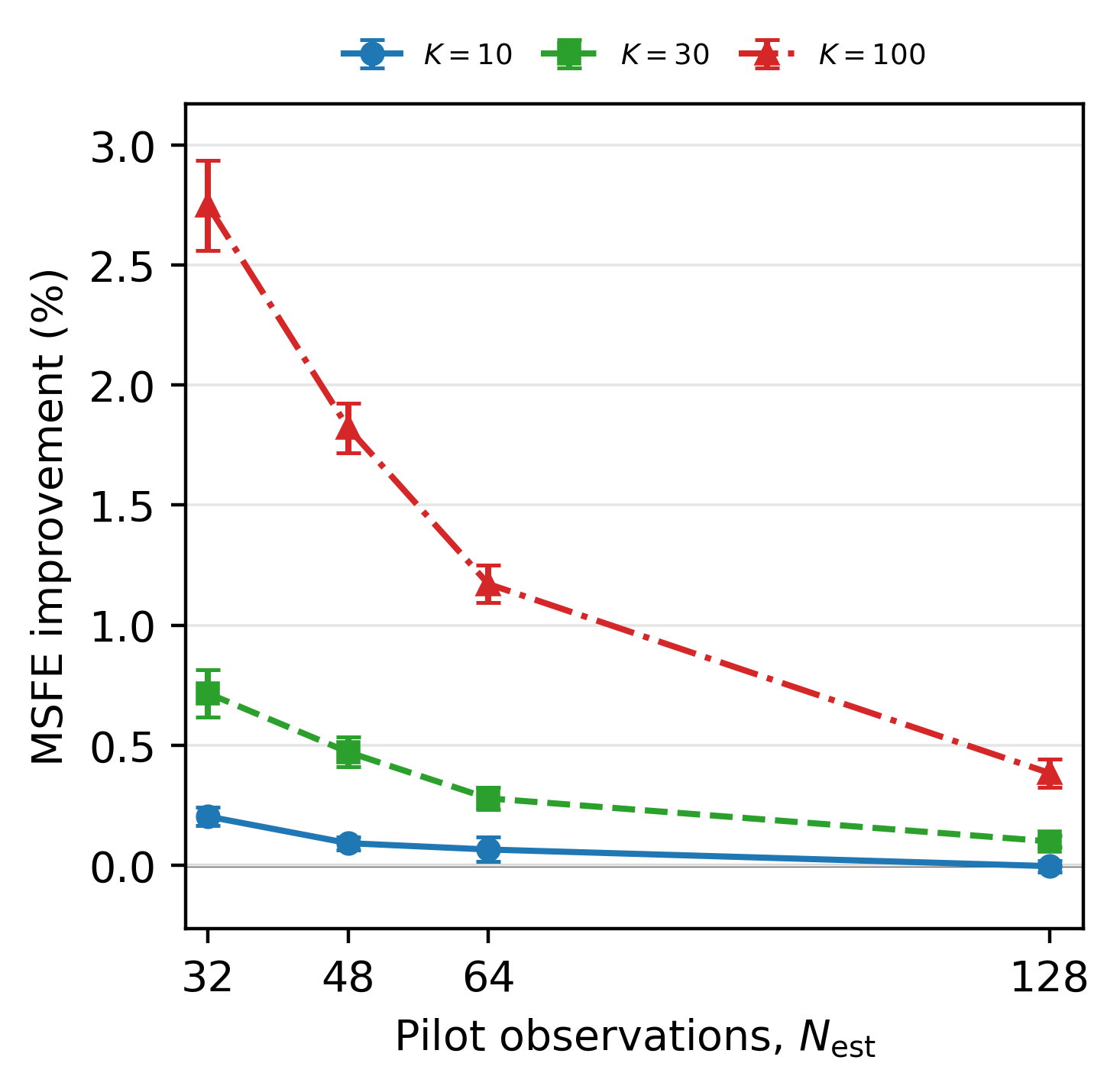}
\caption{Relative \ac{MSFE} improvement of the matched Rician prior over the Gaussian prior as a function of the number of pilot observations $N_{\mathrm{est}}$, for Rician factors $K=10,30,100$. The neural variational estimators are trained for $300{,}000$ updates using the likelihood-derived regularization weight $\zeta=1/(2N_{\mathrm{est}})$. Markers indicate the mean improvement across three paired training seeds, and error bars indicate $95\%$ confidence intervals across seeds. Positive values favor the matched Rician prior.}
\label{fig:vae_improvement}
\end{figure}

\subsection{Sensitivity to the Number of Pilot Observations}
\label{ss:sensitivity}

Fig.~\ref{fig:vae_improvement} compares the relative \ac{MSFE} improvement
obtained with the matched Rician prior for different numbers of pilot
observations. The improvement generally increases as $N_{\mathrm{est}}$
decreases, particularly at higher Rician $K$-factors. For
$N_{\mathrm{est}}=32$, the matched Rician prior reduces the \ac{MSFE} by
approximately $0.20\%$, $0.72\%$, and $2.75\%$ for $K=10$, $30$, and
$100$, respectively. At $N_{\mathrm{est}}=64$, these improvements
decrease to $0.07\%$, $0.28\%$, and $1.17\%$, and at
$N_{\mathrm{est}}=128$, to approximately $0.00\%$, $0.10\%$, and
$0.38\%$. This behavior is consistent with the increased importance
of prior information when fewer pilot observations are available.
The matched Rician prior is therefore most beneficial in the tested
pilot-limited regimes, particularly at higher $K$.

For each operating point and training seed, the neural variational estimators are trained using 100{,}000 training blocks and $300{,}000$ optimization updates. Checkpoints are selected separately for the Gaussian and Rician
estimators by minimizing validation \ac{MSFE} over $5{,}000$ blocks not used
for training. The selected models are evaluated on $10{,}000$ independent
held-out test blocks. Three paired training seeds are used, with Gaussian and Rician estimators sharing the same data within each seed. The validation and test blocks are also shared across seeds. Each block contains $N_{\mathrm{est}}$ pilot observations.

Table~\ref{tab:absolute_msfe_benchmarks} compares the neural
estimators with \ac{LS}, analytical \ac{LMMSE}, the scalar matched
Rician posterior-mean estimator in
\eqref{eq:scalar_rician_posterior_mean}, and the joint
Rician \ac{MMSE} estimator in \eqref{eq:joint_rician_mmse}.
The joint Rician estimator achieves the lowest observed
\ac{MSFE} at all $12$ operating points, followed by the
scalar Rician estimator. The matched Rician \ac{VAE}
achieves lower \ac{MSFE} than \ac{LMMSE} at five operating
points, with the largest improvement of $2.50\%$ at
$N_{\mathrm{est}}=32$ and $K=100$.
At this Rician factor, the improvement decreases to $1.56\%$
and $0.91\%$ for $N_{\mathrm{est}}=48$ and $64$, respectively.
At $N_{\mathrm{est}}=128$, no improvement over \ac{LMMSE}
is observed.

The final column reports the paired \ac{MSFE} difference
$\Delta=\mathrm{MSFE}_{\mathrm{Rician\ VAE}}-
\mathrm{MSFE}_{\mathrm{LMMSE}}$, where negative values favor the
matched Rician \ac{VAE}. The pointwise $95\%$ training-seed
confidence intervals use Student's $t$ distribution with two
degrees of freedom and are conditional on the shared held-out
test blocks.
The intervals exclude zero at five operating points favoring
the matched Rician \ac{VAE} and five favoring \ac{LMMSE}. The remaining
two intervals include zero.
The results show that the matched Rician prior
can improve estimation accuracy over \ac{LMMSE} in pilot-limited
regimes, particularly at high Rician factors. However, both
the scalar Rician posterior-mean estimator and the joint
Rician \ac{MMSE} estimator achieve lower observed \ac{MSFE}
than the matched Rician \ac{VAE} at all $12$ operating points.
The joint estimator gives the lowest observed \ac{MSFE}
throughout the tested conditions.

\begin{figure}[t!]
\centering
\includegraphics[width=\columnwidth]{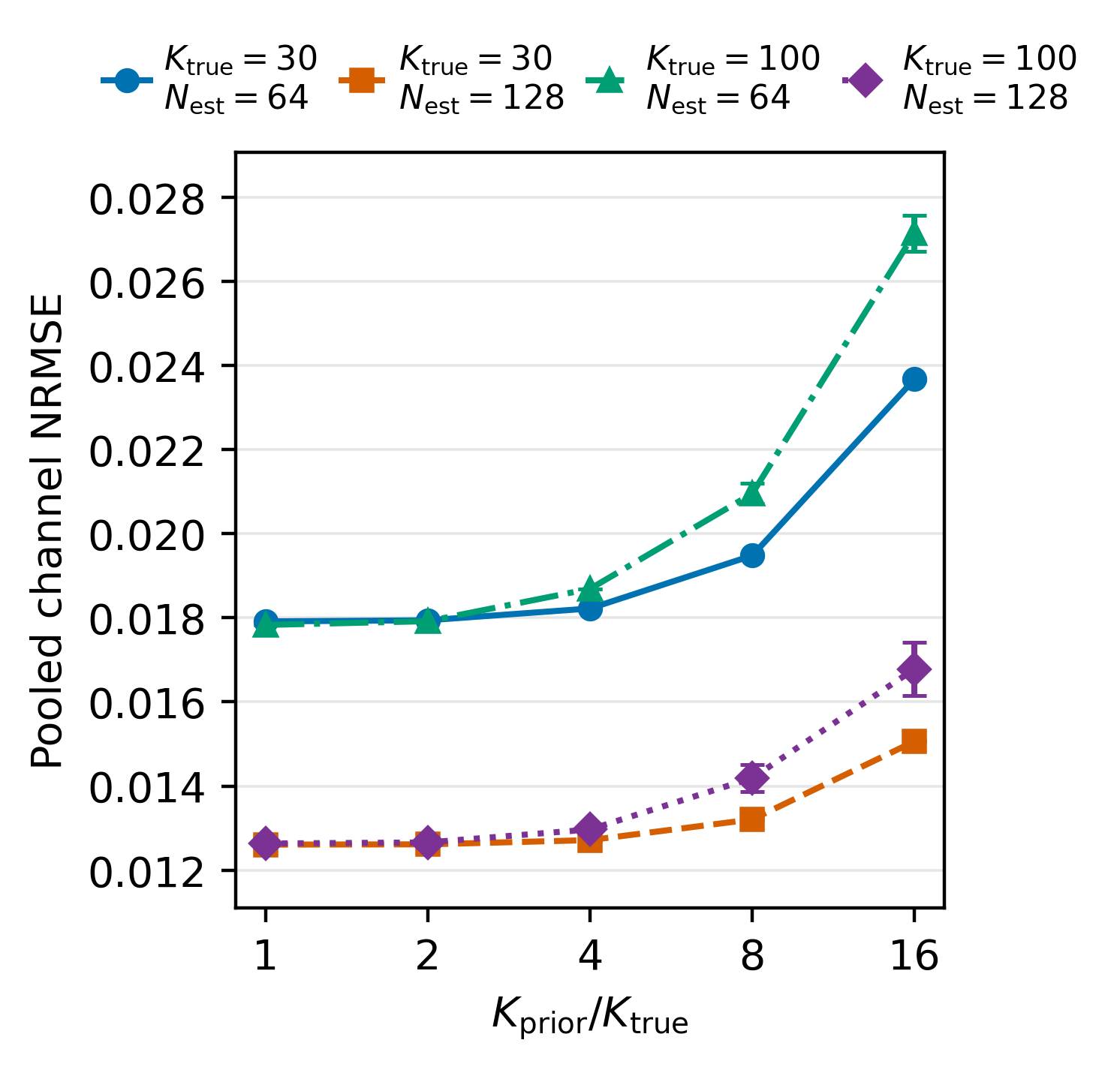}
\caption{Pooled channel \ac{NRMSE} versus
$K_{\mathrm{prior}}/K_{\mathrm{true}}$ for
$K_{\mathrm{true}}\in\{30,100\}$ and
$N_{\mathrm{est}}\in\{64,128\}$.
The neural variational estimators are trained for $300{,}000$
updates using $\zeta=1/(2N_{\mathrm{est}})$.
Markers show mean pooled \ac{NRMSE} across three training seeds,
with error bars indicating $95\%$ confidence intervals.
The matched Rician prior (ratio $1$) gives the lowest
estimation error at all four operating points.
Increasing the assumed Rician factor beyond the
physical value generally increases the error,
particularly at $N_{\mathrm{est}}=64$.}
\label{fig:rician_k_prior_upper_sweep}
\end{figure}

\subsection{Sensitivity to the Assumed Rician Factor}
\label{subsec:kprior_sensitivity}

Fig.~\ref{fig:rician_k_prior_upper_sweep} shows the effect of the
assumed Rician factor on channel \ac{NRMSE} for
$K_{\mathrm{true}}\in\{30,100\}$ and
$N_{\mathrm{est}}\in\{64,128\}$. The physically matched choice
$K_{\mathrm{prior}}=K_{\mathrm{true}}$ gives the lowest \ac{NRMSE}
at all four operating points. Increasing
$K_{\mathrm{prior}}/K_{\mathrm{true}}$ to $2$ produces only a
small change in estimation error, whereas further sharpening
generally increases the \ac{NRMSE}. The increase is particularly
pronounced at $N_{\mathrm{est}}=64$.

With $300{,}000$ training updates and the physically calibrated
regularization weight $\zeta=1/(2N_{\mathrm{est}})$, prior
sharpening does not improve channel estimation over the
matched Rician prior. These results support using the physical
Rician factor under the tested conditions.

\subsection{Non-Gaussian Prior Controls}
\label{subsec:non_gaussian_prior_controls}

\begin{figure}[b!]
\centering
\includegraphics[width=\columnwidth]{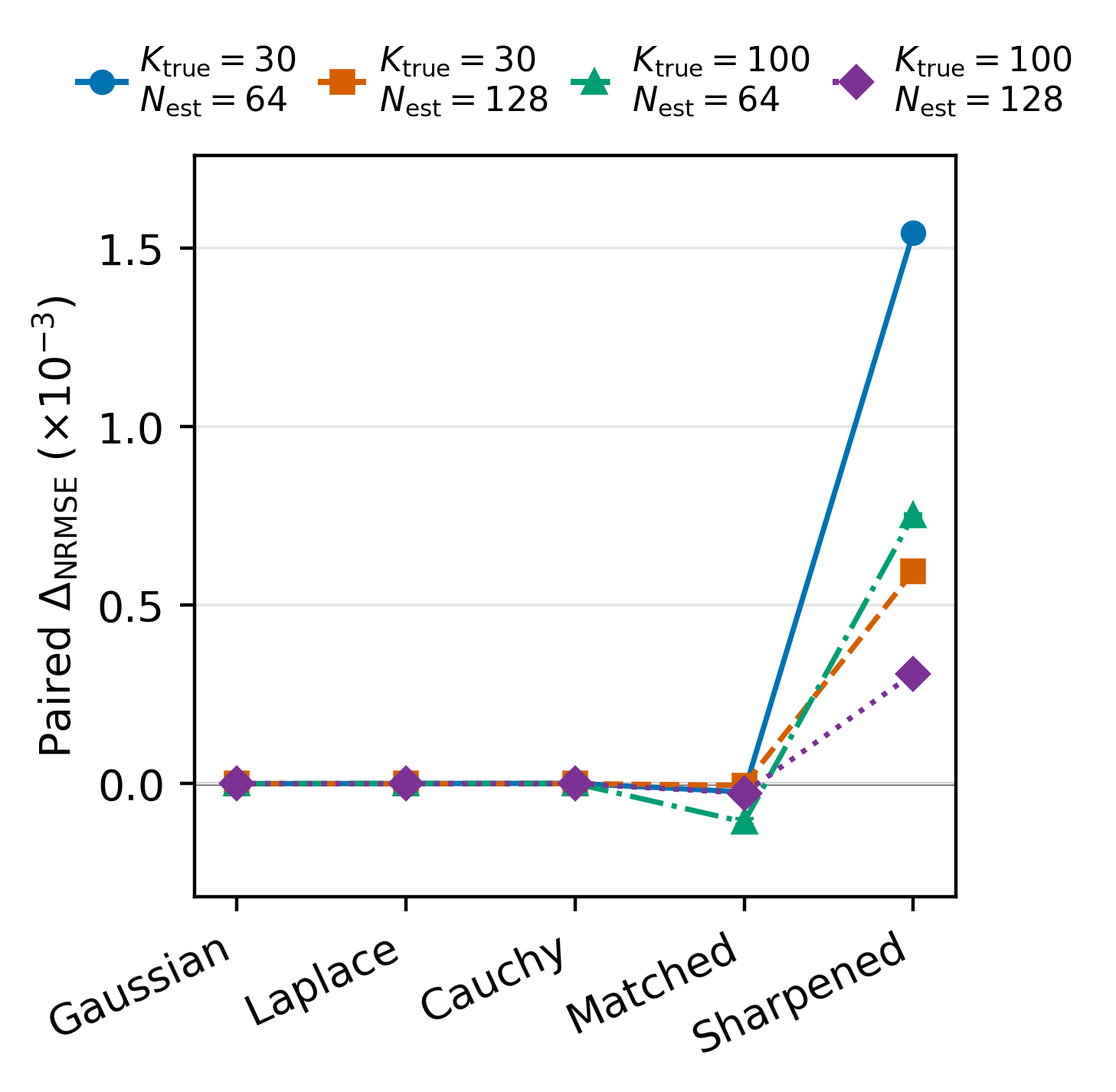}
\caption{Paired channel \ac{NRMSE} differences relative to the Gaussian
prior for Laplace, Cauchy, matched Rician, and sharpened Rician
priors, with $K_{\mathrm{true}}\in\{30,100\}$ and
$N_{\mathrm{est}}\in\{64,128\}$. The neural variational
estimators are trained for $300{,}000$ updates using
$\zeta=1/(2N_{\mathrm{est}})$. Negative differences favor the
tested prior. Laplace and Cauchy perform similarly to Gaussian,
whereas the matched Rician prior gives the lowest \ac{NRMSE} at all
four operating points. Prior sharpening increases the error
relative to Gaussian and the matched Rician prior.}
\label{fig:non_gaussian_prior_control}
\end{figure}

Fig.~\ref{fig:non_gaussian_prior_control} compares Gaussian, Laplace,
Cauchy, matched Rician, and sharpened Rician priors for
$K_{\mathrm{true}}\in\{30,100\}$ and
$N_{\mathrm{est}}\in\{64,128\}$. We observe that the Laplace and
Cauchy priors give \ac{NRMSE} values close to the Gaussian baseline,
whereas the physically matched Rician prior gives the lowest
estimation error at all four operating points. In contrast,
the sharpened Rician prior increases \ac{NRMSE} relative to the
matched prior, particularly at $N_{\mathrm{est}}=64$.

These results indicate that the matched Rician advantage is not
reproduced by the two tested heavy-tailed priors. The improvement
is consistent with the use of channel-specific Rician prior
structure rather than non-Gaussian regularization alone.
Under the tested conditions, additional prior sharpening
does not improve channel estimation.

\section{Discussion}
\label{sec:disc}

\subsection{Latent Prior Mismatch and Regularization}

The results show that physical latent-prior mismatch matters
most in pilot-limited regimes with strong Rician components.
At $K=0$, the Gaussian and Rician priors coincide exactly.
For the tested positive Rician factors, the matched-prior
advantage increases with $K$ and generally decreases with
$N_{\mathrm{est}}$. These trends do not establish a universal
threshold for either parameter. Additional Rician prior
sharpening increases estimation error, while the tested
Laplace and Cauchy controls do not reproduce the
matched Rician gains.

The joint Rician \ac{MMSE} estimator achieves lower observed
\ac{MSFE} than the scalar Rician posterior-mean estimator
at all $12$ operating points. The joint estimator uses
the complete \ac{LS} noise covariance and therefore
accounts for correlations between estimation errors
associated with different transmitter columns. The paired
95\% confidence intervals exclude zero at $6$ of the
$12$ operating points, indicating that the additional
improvement over scalar Rician estimation is modest
under the tested conditions.

Both analytical Rician estimators achieve lower observed
\ac{MSFE} than the Gaussian and matched Rician variational
estimators at all $12$ operating points. The joint Rician
\ac{MMSE} estimator provides the model-matched Bayesian
reference, with all $24$ paired block-bootstrap confidence
intervals favoring joint estimation over the two neural
estimators. The remaining neural estimation gap may reflect
posterior approximation, amortization, or optimization
errors. The experiments do not separate these contributions. The numerical endpoints of the paired bootstrap confidence intervals are provided in the supplementary material.

\vspace{-3mm}
\subsection{Receiver Interpretation and Limitations}

The experiments consider receiver-side estimation of a
block-static $2\times2$ complex channel with independently
generated Rician coefficients and a fixed physical
latent-to-channel mapping. The matched prior is exact
under this controlled model, but general polarization
channels may exhibit dependence between coefficients.
Temporal fading correlation, polarization dynamics,
pointing variations, and hardware drift are not modeled.

The channel model further assumes negligible or previously
compensated propagation delays. Resolvable multipath
would require a frequency-selective matrix channel and
a higher-dimensional latent representation.

In \ac{CVQKD}, channel estimation uncertainty contributes to
uncertainty in estimated transmission and noise parameters
used in secret-key-rate calculations. Consequently, even
modest improvements in estimation accuracy may affect
protocol performance. However, the reported \ac{NRMSE}
and \ac{MSFE} improvements do not establish increased
secret-key rates or composable security. Such conclusions
require propagation of channel estimation error and
uncertainty through protocol-level parameter estimation
and security analysis.

Finally, the matched Rician experiments assume known
channel statistics. Practical implementations may require
joint estimation or tracking of the channel state and its
prior parameters, particularly under time-varying
propagation conditions.

\subsection{Complexity and Implementation}
\label{ss:complexity}

For the fixed two-polarization model, forming
$\bm S_{yx}=\bm Y\bm X^H$ and
$\bm S_{xx}=\bm X\bm X^H$ requires
$O(N_{\mathrm{est}})$ operations. The resulting
fixed-dimensional statistics provide a 16-dimensional real
encoder input. The encoder has two hidden layers of 424
units, an eight-dimensional latent space, and
$P_{\phi}=194,208$ trainable parameters. Each forward pass
requires 193,344 dense multiply--accumulate operations,
followed by the fixed latent-to-channel mapping without
a learned decoder.

The \ac{LS} and analytical \ac{LMMSE} estimators require
$O(N_{\mathrm{est}})$ operations and a fixed-size linear
solve. The scalar Rician posterior-mean estimator
additionally evaluates modified Bessel functions for
each channel coefficient. The joint Rician \ac{MMSE}
estimator retains the complete \ac{LS} noise covariance
and evaluates a one-dimensional phase integral using
$Q$ quadrature nodes. Analytical integration of the
Gaussian component and one Rician phase reduces the
quadrature complexity from $O(Q^2)$ to $O(Q)$.
For fixed polarization dimension, the resulting
per-block computational complexity is
$O(N_{\mathrm{est}}+Q)$.

We evaluate the joint posterior mean using $Q=256$
quadrature nodes and verify convergence against $Q=512$.
All $120,000$ held-out blocks satisfy the numerical
convergence criterion without requiring additional
quadrature nodes.

Neural inference additionally requires feature
standardization and an encoder forward pass, giving
$O(N_{\mathrm{est}}+P_{\phi})$ complexity. All tested neural
priors have identical inference complexity because the
prior densities are evaluated only during training.
Table~\ref{tab:estimator_complexity} summarizes these costs.
The Gaussian \ac{KL} term is analytical, whereas Rician,
Laplace, and Cauchy priors use sampled forward \ac{KL}
evaluation over fixed-dimensional latent variables.

\begin{table}[h!]
\vspace{-2mm}
\centering
\caption{Computational complexity for a fixed $2\times2$
complex channel, where $P_{\phi}=194,208$.}
\label{tab:estimator_complexity}
\small
\setlength{\tabcolsep}{12pt}
\begin{tabular}{lcc}
\hline
Estimator & Parameters & Complexity \\
\hline
LS & 0 & $O(N_{\mathrm{est}})$ \\
Analytical LMMSE & 0 & $O(N_{\mathrm{est}})$ \\
Scalar Rician & 0 & $O(N_{\mathrm{est}})$ \\
Joint Rician MMSE & 0 & $O(N_{\mathrm{est}}+Q)$ \\
Gaussian VAE & $P_{\phi}$ &
$O(N_{\mathrm{est}}+P_{\phi})$ \\
Matched Rician VAE & $P_{\phi}$ &
$O(N_{\mathrm{est}}+P_{\phi})$ \\
\hline
\end{tabular}
\end{table}

\section{Conclusion}
\label{sec:conclusion}

This paper examined physical latent-prior mismatch in
variational dual-polarization quantum \ac{SATCOM} channel
estimation. The eight-dimensional latent representation
corresponds directly to the complex channel coefficients,
with a fixed physical latent-to-channel mapping. Holding
the posterior family, forward \ac{KL} objective, observation
model, and estimator architecture fixed provides a
controlled comparison of Gaussian, matched Rician, and
alternative latent priors.

The Rician and Gaussian priors coincide at the Rayleigh
limit. For positive Rician factors, the matched Rician
prior improves channel \ac{NRMSE} over the Gaussian prior,
particularly at high $K$ and small $N_{\mathrm{est}}$.
Additional Rician prior sharpening increases estimation
error, while the tested Laplace and Cauchy controls do
not reproduce the matched Rician gains. At $K=100$ and
$N_{\mathrm{est}}=32$, the matched Rician \ac{VAE} reduces
\ac{MSFE} relative to \ac{LMMSE} by $2.50\%$.
The joint Rician \ac{MMSE} estimator, which accounts for
the complete \ac{LS} noise covariance, achieves a
$3.33\%$ reduction relative to \ac{LMMSE} at the same
operating point. It gives lower observed \ac{MSFE} than
the scalar Rician posterior-mean estimator and both
neural estimators at all 12 tested operating points.
The comparison establishes the remaining estimation
gap between the variational estimators and the
model-matched Bayesian posterior mean.

The results concern receiver-side estimation under a
controlled channel law, not \ac{CVQKD} security.
Future work can examine correlated and time-varying
polarization channels, adaptive prior estimation, and
the propagation of channel estimation uncertainty into
quantum communication performance.

\section*{Acknowledgment}
This work was supported by the Research Council of Norway. 
OpenAI ChatGPT was used for language editing. The authors reviewed and verified all content.

\bibliographystyle{IEEEtran}
\bibliography{IEEEabrv,prior}

\end{document}